\documentclass[11pt]{article}

\usepackage[utf8]{inputenc}
\usepackage[T1]{fontenc}
\usepackage{lmodern}
\usepackage[margin=1in]{geometry}
\usepackage{amsmath,amssymb,amsthm}
\usepackage{mathtools}
\usepackage{booktabs}
\usepackage{enumitem}
\usepackage{xcolor}
\usepackage{listings}
\usepackage{tikz}
\usetikzlibrary{positioning,fit,arrows.meta}
\usepackage[colorlinks=true,linkcolor=blue!50!black,citecolor=blue!50!black,urlcolor=blue!50!black]{hyperref}

\newtheorem{theorem}{Theorem}[section]
\newtheorem{lemma}[theorem]{Lemma}
\newtheorem{proposition}[theorem]{Proposition}
\newtheorem{corollary}[theorem]{Corollary}
\theoremstyle{definition}
\newtheorem{definition}[theorem]{Definition}

\theoremstyle{remark}
\newtheorem{remark}[theorem]{Remark}

\newcommand{\Q}{\mathbb{Q}}
\newcommand{\tw}{\operatorname{tw}}
\newcommand{\Prim}{\operatorname{Pr}}
\newcommand{\poly}{\operatorname{poly}}

\title{Limits and Constants of OpenAI's Claimed\\ETH Refutation}
\author{Fabio Dias\thanks{Corresponding author:
  \href{mailto:f.dias@surrey.ac.uk}{\texttt{f.dias@surrey.ac.uk}}}\\ University of Surrey
  \and Yvens Lima\\ Universidade Federal do Cear\'a}
\date{October 2026}

\begin{document}
\maketitle

\begin{abstract}
A recent OpenAI preprint claims a deterministic algorithm that counts the satisfying assignments of a $k$-CNF formula in time $2^{O_k(n\log\log n/\sqrt{\log n})}$, and $2^{O(n/\sqrt{\log n})}$ at bounded clause density, which would refute the Exponential Time Hypothesis. The headline bound is proved by tensor contraction; the body develops an interpolation algorithm with the weaker exponent $n(\log\log n)^2/\sqrt{\log n}$. This note does four things. First, it explains the interpolation algorithm in plain terms. Second, it asks how large $n$ must be for the bounds to bite. The preprint states no constants, so we treat the hidden constant $C$ as a parameter: the general bound beats the $2^n$ of brute force only beyond $n\approx 10^{32}$ for $C=2$, $10^{126}$ for $C=3$ and $10^{3517}$ for $C=10$. We then extract the constants from the tensor-contraction proof: for every input with $m$ clauses at bounded density, the running time it certifies stays above $2^n$ unless $\log_2(n+m)>10^9$, and on random 3-CNF at the satisfiability threshold unless $\log_2 n>5\times10^{11}$. So the bounds are subexponential as $n\to\infty$, but the proof certifies no speed-up over brute force for any input of bounded density that could ever be written down. Third, for the clause-product form of the interpolation algorithm we prove limits in a cost model matching the preprint's evaluation. By a short algebraic lemma, polynomial cost needs components of $O(\log^3 n)$ variables with per-point evaluation, or primal treewidth $O(\log^2 n)$ with dynamic programming over the groups. So the scheme costs $2^{\Omega(\sqrt{\mathrm{tw}})}$ at treewidth $\mathrm{tw}$, $2^{\Omega(\sqrt n)}$ with high probability on random 3-CNF above clause density 3, and $2^{\Omega(n^{1/3})}$ even on trivially countable path formulas. We explain why the preprint's sharper routes escape these limits. Fourth, we report experiments.
\end{abstract}

\section{Introduction}
\enlargethispage{2pt}

The Exponential Time Hypothesis (ETH) of Impagliazzo and Paturi~\cite{IP01} states that 3-SAT
on $n$ variables needs time $2^{\delta n}$ for some $\delta>0$. Together with the sparsification
lemma~\cite{IPZ01} it underlies a large part of fine-grained complexity~\cite{VW18}. On
17~October 2026 OpenAI posted a preprint~\cite{openai}, revised on 25~October, whose Theorem~1.1
states that $\#k$-SAT can be solved deterministically in time
$(n+m)^{O_k(1)}2^{O_k(n\log\log n/\sqrt{\log n})}$ and polynomial space, and in time
$(n+m)^{O_{k,D}(1)}2^{O_{k,D}(n/\sqrt{\log n})}$ when $m\le Dn$. This refutes ETH and its counting
version \#ETH~\cite{DHMTW14}. The preprint uses natural logarithms throughout. Its Appendix~D proves
both bounds by contracting Boolean tensor networks. The body of the preprint develops an
\emph{interpolation algorithm} (its Proposition~6.4) with the weaker exponent
\[
\rho(n)=\frac{n(\log\log n)^2}{\sqrt{\log n}}.
\]
At the time of writing the claim is under independent review. We take the preprint's proofs at face
value and do not assess their correctness.

The interpolation algorithm comes in two forms. The \emph{clause-product} form (Sections~2.1--2.8
of the preprint) multiplies clause polynomials of a formula in which every variable occurs at most
three times, reached by sparsification and variable copies. For general $k$-CNF the preprint
derives only $2^{O_k(n(\log\log n)^{k+1}/(\log n)^{1/(2k)})}$ from it (its Section~2.8), but on
formulas of bounded density, where variable copies alone give $O(n)$ variables, it yields
$2^{O(\rho(n))}$. The \emph{parity} form (Section~2.9 and Theorem~2.2) replaces clauses by
sparse polynomial equations over $\mathbb{F}_2$ and achieves $2^{O(\rho(n))}$ in general. Our
contributions are as follows.

\begin{enumerate}[leftmargin=*]
\item \textbf{A plain-language account} (Section~\ref{sec:explain}). We explain the interpolation
  algorithm step by step, with a three-variable example in which every quantity can be checked by
  hand, and we say which parts of the sixty-page proof carry the weight and which are engineering.
\item \textbf{The constants} (Section~\ref{sec:constants} and Appendix~\ref{app:crossover}). The
  preprint states no constants. We tabulate where its bounds beat brute force as a function of the
  hidden constant, and we extract the constants of the tensor-contraction proof from its
  Appendix~D. For every input of bounded density, the running time that proof certifies is above
  $2^n$ unless $\log_2(n+m)>1.03\times 10^9$.
\item \textbf{A vanishing-product lemma} (Lemma~\ref{lem:vanish}). In the clause-product scheme, each
  occurrence of a variable carries a polynomial ``own coefficient'' that equals $1$ at $T=0$; their
  product has degree at most the degree up to which the recovery is exact, so it cannot vanish to
  high order at every evaluation point. Hence at some point at most $q$ occurrences lose their own
  variable, where $q$ is the Taylor order used there.
\item \textbf{Necessary conditions for polynomial cost} (Theorems~\ref{thm:structure}
  and~\ref{thm:necessary}), in a cost model that charges what the preprint's evaluation pays
  (Section~\ref{sec:framework}). At that point the formula's primal treewidth is at most
  $(w+1)s-1+q$, where $s$ is the group size and $w$ the treewidth of the groups' interaction graph.
  Polynomial cost therefore forces primal treewidth $O(\log^2 n)$, and with the preprint's
  per-point evaluation and bounded occurrence it forces the formula to be a union of
  $O(\log^3 n)$-variable pieces.
\item \textbf{Barriers} (Section~\ref{sec:barriers}). In the same model, the clause-product scheme
  costs $2^{\Omega(\sqrt{\tw})}$ on formulas of treewidth $\tw$; $2^{\Omega(\sqrt n)}$ w.h.p.\ on
  random 3-CNF with $m=\Delta n$, $\Delta>3$, both after the preprint's reduction to bounded
  occurrence and after its sparsification step; and $2^{\Omega(n^{1/3})}$ with the preprint's
  evaluation even on path formulas.
\item \textbf{Experiments} (Section~\ref{sec:experiments}) and code (Appendix~\ref{app:code}).
\end{enumerate}

These results say nothing against the preprint's subexponential bounds. The limits concern the
clause-product scheme only. The parity form and the tensor contraction fall outside them, and we
explain why (Section~\ref{sec:scope}). For the clause-product scheme on expanding inputs of bounded
density, the cost lies between $2^{\Omega(\sqrt n)}$ and $2^{O(\rho(n))}$, and whether it is near
either end is open (Section~\ref{sec:gap}).

\section{Related work}\label{sec:related}

\paragraph{ETH, SETH and exact algorithms.} ETH and the Strong ETH were formulated by
Impagliazzo and Paturi~\cite{IP01}; sparsification~\cite{IPZ01} with its quantitative form by
Calabro, Impagliazzo and Paturi~\cite{CIP06,Cal09} transfers ETH to many problems, and its
counting-preserving form~\cite{DHMTW14} gives \#ETH. Exponential algorithms for $k$-SAT include
randomised local search~\cite{Sch99} and PPSZ~\cite{PPSZ05}, with Hertli's analysis~\cite{Her14},
biased PPSZ~\cite{HKZZ19} and Scheder's improved analysis~\cite{Sch24}. For counting, Dahll\"of,
Jonsson and Wahlstr\"om~\cite{DJW05} and Wahlstr\"om~\cite{Wah08} gave exponential-time bounds for
\#2-SAT and \#3-SAT, and Kutzkov~\cite{Kut26} recently improved the bound for \#3-SAT. Fine-grained
complexity is surveyed in~\cite{VW18}.

\paragraph{Algebraic and interpolation methods.} Counting through polynomial identities goes back
to Valiant's permanent~\cite{Val79}; inclusion--exclusion~\cite{BHK09}, algebraic sieving for
paths and packings~\cite{Kou08,Wil09} and determinant sums~\cite{Bjo10} made it a design tool. The
polynomial method of Williams~\cite{Wil14} and its descendants~\cite{AWY15} give
super-polylogarithmic savings for many problems; for systems of polynomial equations over
$\mathbb{F}_2$ it yields $2^{(1-\epsilon)n}$-type algorithms~\cite{LPTWY17,BKW19,Din21}.
Curticapean's block interpolation~\cite{Cur18} is a counting framework for tight lower bounds. The
preprint~\cite{openai} combines a one-parameter deformation of clause inputs with Hermite
interpolation, expander-code parity checks and sparsification (we describe it in
Section~\ref{sec:explain}), and proves its sharpest bounds by a recursive contraction of Boolean
tensor networks with two-adic Hermite recovery.

\paragraph{Treewidth and counting.} Model counting is fixed-parameter tractable in the treewidth
of the primal or incidence graph~\cite{FMR08,SS10}; see~\cite{Bod93} for treewidth in general and
\cite{BK10,BK11,GD04} for the upper- and lower-bound heuristics used in our experiments; exact
counters based on component caching, such as sharpSAT~\cite{Thu06} and d4~\cite{LM17}, are the
practical standard. Random graphs above the giant-component threshold have linear
treewidth~\cite{Gao12,LLO12}, and linear width of random CNF is a standard consequence of their
expansion; we include a short self-contained proof of the form we use (Lemma~\ref{lem:randtw}).

\paragraph{Barriers for methods.} Our results are barriers for a method, not for a problem, in
the spirit of the limits of the laser method for matrix
multiplication~\cite{AFLG15,AW18,Alm19,CVZ19}. In proof complexity, expansion is the common source
of lower bounds for resolution~\cite{CS88,BW01} and the polynomial calculus~\cite{BI10,AR03}; our
treewidth condition plays the same role for the clause-product scheme. Restricted-model lower bounds also
underlie the low-degree method for average-case problems~\cite{Hop18,KWB19}, whose limits have
been probed by counterexamples built from decodable codes~\cite{HW21,BHJK25}.

\paragraph{Benchmarks.} The SATLIB uniform random 3-SAT families~\cite{HS00} sample the
satisfiability threshold near $m/n\approx 4.26$~\cite{MSL92,CA96}.

\section{The OpenAI argument in plain terms}\label{sec:explain}

This section is for readers who know what a satisfiability problem is and little else. It uses
no algebra beyond school-level polynomials. It describes the preprint's interpolation algorithm; the
small example is ours. Specialists can go straight to Section~\ref{sec:explain-precise}, which
restates it in the preprint's terms.

\subsection{The problem: counting good settings}

Picture $n$ on/off switches and a list of rules, each mentioning at most three switches, such as
``switch~1 is on, or switch~2 is off, or switch~3 is on''. A setting of all the switches is
\emph{good} if it obeys every rule. The task is to count the good settings; counting is at least as
hard as deciding whether one exists.

The obvious method tries every setting. There are $2^n$ of them, which for 100 switches is about
$10^{30}$, far beyond any computer. The Exponential Time Hypothesis is the belief that nothing
fundamentally better is possible: that the running time must grow like $2^{\delta n}$ for some
fixed $\delta>0$, as if a fixed fraction of the switches always had to be tried exhaustively. The
preprint claims an algorithm whose running time grows more slowly than that.

\subsection{Why splitting helps}

Suppose the switches fell into two teams of 50, and every rule mentioned switches of one team
only. Then we could count the good settings of the first team (at most $2^{50}$ to try), count those
of the second, and multiply, because every good setting of one team combines with every good
setting of the other. That is $2\times 2^{50}$ settings to try instead of $2^{100}$, about a million
billion times fewer. More, smaller teams would be better still.

Hard puzzles typically do not split like this. In a random puzzle the rules tie the switches into a
well-knit net: however the switches are divided into teams, a large share of the rules mention
switches of two different teams. Graphs of this kind are called \emph{expanders}.

\subsection{The idea: a dial that rewires the rules}

First, a way to count. Give each rule a score for each setting: $1$ if the setting obeys it, $0$ if
not. Multiply the scores of all the rules, which gives $1$ for a good setting and $0$ for any other,
and add up over all settings. The total is the number of good settings. We call this the
\emph{counting recipe}.

The preprint does not split the puzzle. It builds a whole family of related puzzles, one for each
position of a dial $T$, with three properties.
\begin{enumerate}[leftmargin=*]
\item At $T=0$ the puzzle is the original one, so the recipe gives the count.
\item At certain other dial positions the puzzle \emph{does} split into small teams, apart from a
  little controlled leakage, described in the next subsection.
\item The number that the recipe produces depends on the dial in a very regular way: as a
  polynomial in $T$, the kind of curve that can be reconstructed from a few readings.
\end{enumerate}

The rewiring works like this. A rule does not read its switches directly; it reads blends of
switches whose mix depends on the dial. In one rule, ``switch~2'' might become
``$(1-T)$ times switch~2, plus $T$ times switch~3''. At $T=0$ this is switch~2 itself. At $T=1$ it
is switch~3, and that rule no longer mentions switch~2 at all, so the link the rule made through
switch~2 is gone. Different rules can be rewired differently.

At most dial positions the blends are not simply 0 or 1, so the scores are not either, and the
number the recipe produces there does not count anything. It is just a number. That does not matter, because all we need are
points on the curve.

\subsection{Leakage through summaries}

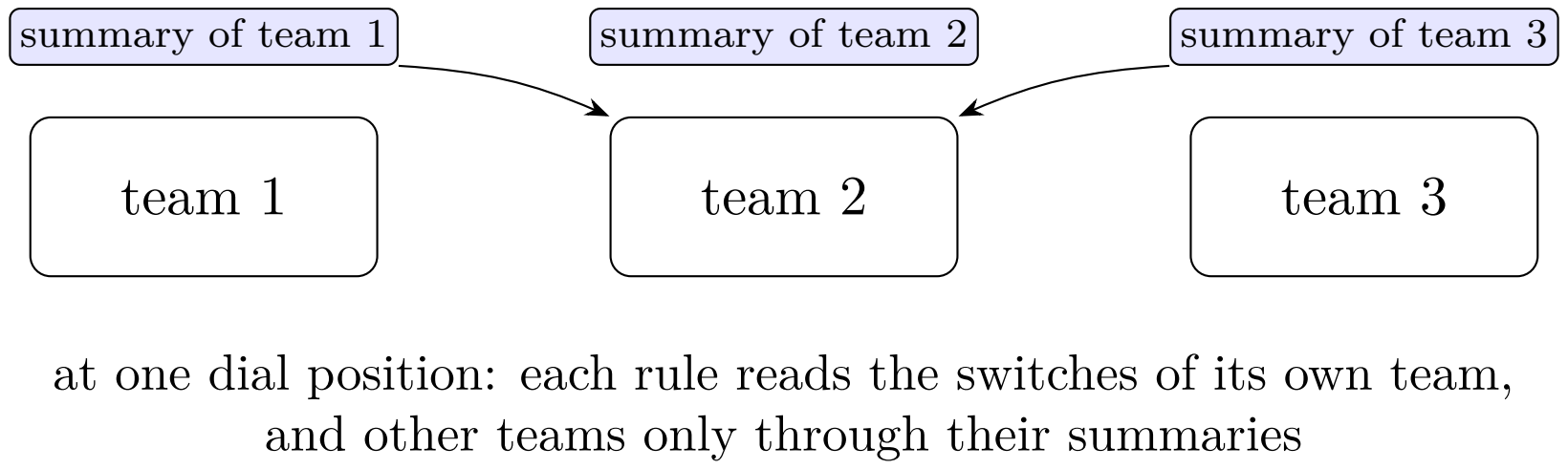
\begin{figure}[t]
\centering
\begin{tikzpicture}[group/.style={draw,rounded corners,minimum width=2.4cm,minimum height=1.1cm},
  dial/.style={draw,rounded corners=2pt,inner sep=2pt,fill=blue!10,font=\scriptsize},>=Stealth]
\node[group] (g1) {team 1};
\node[group,right=1.6cm of g1] (g2) {team 2};
\node[group,right=1.6cm of g2] (g3) {team 3};
\node[dial,above=0.35cm of g1] (y1) {summary of team 1};
\node[dial,above=0.35cm of g2] (y2) {summary of team 2};
\node[dial,above=0.35cm of g3] (y3) {summary of team 3};
\draw[->] (y1.south east) to[bend left=10] (g2.north west);
\draw[->] (y3.south west) to[bend right=10] (g2.north east);
\node[below=0.4cm of g2,align=center,font=\small]
  {at one dial position: each rule reads the switches of its own team,\\
   and other teams only through their summaries};
\end{tikzpicture}
\caption{The split at one dial position. Fixing the value of every summary makes the teams
independent, so each team can be counted on its own; the price is the number of possible
combinations of summary values.}
\label{fig:design}
\end{figure}

At a single dial position it is impossible to cut every link of a hard puzzle. The preprint lets
each team publish a few \emph{summaries} of its switches, such as ``how many switches of team~3
are on'', and lets the rules of other teams read those summaries instead of the switches
themselves (Figure~\ref{fig:design}). A summary takes only a few possible values, between 0 and 10
say.

To produce the number at that dial position, go through every combination of summary values. For
each combination the teams are independent again, so count each team on its own and multiply.
The price is the number of combinations of summary values that have to be gone through.

\subsection{Reading the answer off the curve}

The number we want is the value of the curve at $T=0$, where nothing splits. We can compute the
curve only at the special dial positions where the puzzle does split. But a polynomial is rigid.
A straight line is fixed by two of its points; a curve of degree $d$ by $d+1$ points; and fewer
points will do if at each one we also know the slope, the rate of change of the slope, and so on.
So the recipe is: compute the curve at enough splitting positions, reconstruct it, and read off its
value at $T=0$. That value is the exact count.

The degree of the curve grows with the number of switches and with how strongly the blends depend
on the dial. To keep the number of dial positions small, the preprint uses slopes and higher
rates of change at each position. These are the expensive part. The split holds exactly at the
chosen position, not slightly away from it, so the rates of change need a separate trick, whose
cost grows quickly with how many of them are wanted.

\subsection{A worked example}\label{ex:toy}

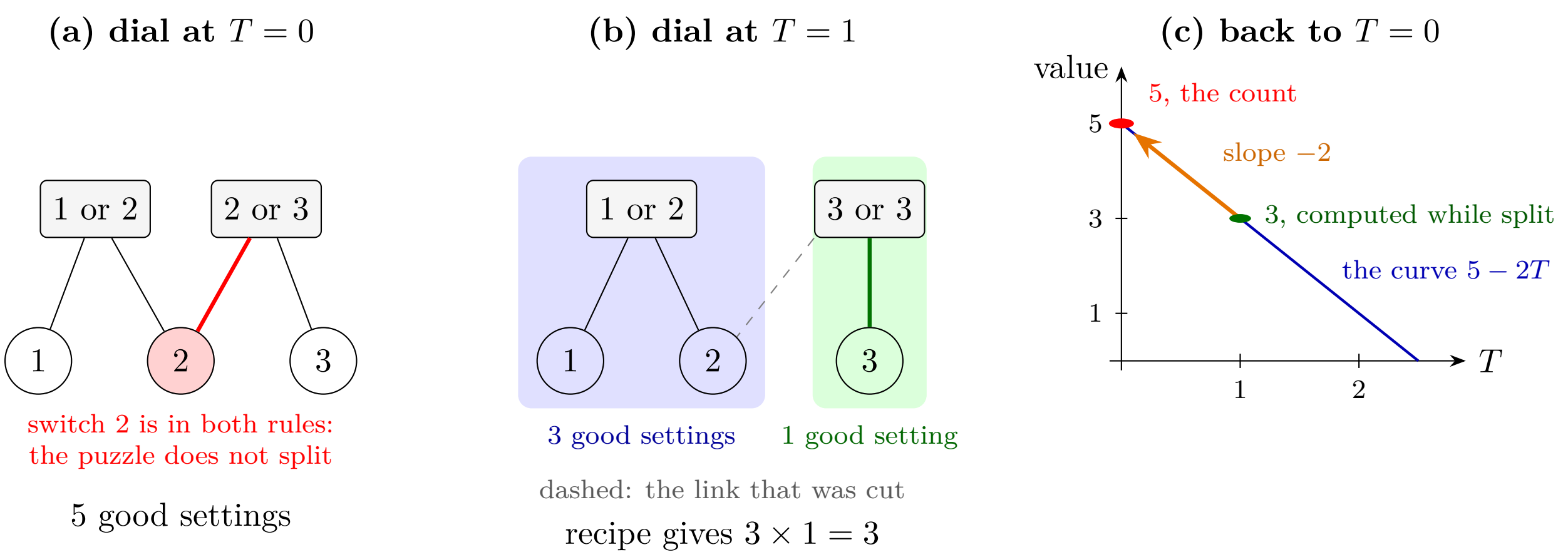
\begin{figure}[t]
\centering
\resizebox{\textwidth}{!}{%
\begin{tikzpicture}[
  var/.style={draw,circle,minimum size=7mm,inner sep=0pt,font=\small},
  cls/.style={draw,rectangle,rounded corners=2pt,minimum width=11mm,minimum height=6mm,font=\small,fill=gray!8},
  >=Stealth]
\begin{scope}
\node[font=\small\bfseries] at (1.5,3.45) {(a) dial at $T=0$};
\node[var] (a1) at (0,0) {$1$};
\node[var,fill=red!18] (a2) at (1.5,0) {$2$};
\node[var] (a3) at (3,0) {$3$};
\node[cls] (aC1) at (0.6,1.6) {$1$ or $2$};
\node[cls] (aC2) at (2.4,1.6) {$2$ or $3$};
\draw (aC1) -- (a1); \draw (aC1) -- (a2);
\draw[red,very thick] (aC2) -- (a2);
\draw (aC2) -- (a3);
\node[font=\scriptsize,red,align=center] at (1.5,-0.85) {switch 2 is in both rules:\\ the puzzle does not split};
\node[font=\small] at (1.5,-1.65) {5 good settings};
\end{scope}
\begin{scope}[xshift=5.6cm]
\node[font=\small\bfseries] at (1.6,3.45) {(b) dial at $T=1$};
\fill[blue!12,rounded corners] (-0.55,-0.5) rectangle (2.05,2.15);
\fill[green!14,rounded corners] (2.55,-0.5) rectangle (3.75,2.15);
\node[var] (b1) at (0,0) {$1$};
\node[var] (b2) at (1.5,0) {$2$};
\node[var] (b3) at (3.15,0) {$3$};
\node[cls] (bC1) at (0.75,1.6) {$1$ or $2$};
\node[cls] (bC2) at (3.15,1.6) {$3$ or $3$};
\draw (bC1) -- (b1); \draw (bC1) -- (b2);
\draw[gray,dashed] (bC2.south west) -- (b2.north east);
\draw[green!45!black,very thick] (bC2) -- (b3);
\node[font=\scriptsize,blue!60!black] at (0.75,-0.8) {3 good settings};
\node[font=\scriptsize,green!40!black] at (3.15,-0.8) {1 good setting};
\node[font=\scriptsize,gray!70!black,align=center] at (1.6,-1.35) {dashed: the link that was cut};
\node[font=\small] at (1.6,-1.85) {recipe gives $3\times 1=3$};
\end{scope}
\begin{scope}[xshift=11.4cm,xscale=1.25,yscale=0.5]
\node[font=\small\bfseries] at (1.5,6.9) {(c) back to $T=0$};
\draw[->] (-0.1,0) -- (2.9,0) node[right,font=\small] {$T$};
\draw[->] (0,-0.2) -- (0,6.2) node[left,font=\small] {value};
\foreach \t in {1,2} \draw (\t,0.15) -- (\t,-0.15) node[below,font=\scriptsize] {\t};
\foreach \h in {1,3,5} \draw (0.05,\h) -- (-0.05,\h) node[left,font=\scriptsize] {\h};
\draw[blue!70!black,thick] (0,5) -- (2.5,0);
\node[font=\scriptsize,blue!70!black,right] at (1.75,1.9) {the curve $5-2T$};
\draw[orange!90!black,very thick,->] (1,3) -- (0.1,4.8);
\node[font=\scriptsize,orange!80!black,right] at (0.75,4.35) {slope $-2$};
\fill[green!45!black] (1,3) circle (2.6pt);
\node[font=\scriptsize,green!35!black,right] at (1.1,3.05) {3, computed while split};
\fill[red] (0,5) circle (3pt);
\node[font=\scriptsize,red,right] at (0.12,5.6) {5, the count};
\end{scope}
\end{tikzpicture}}
\caption{The worked example. Circles are switches, boxes are rules. (a) With the dial at $0$ the
two rules share switch~2. (b) With the dial at $1$ the second rule has been rewired to read
switch~3 instead, so the puzzle splits into the blue team $\{1,2\}$ and the green team $\{3\}$, and
the recipe's number is a product of two small counts. (c) The recipe's number is the straight line
$5-2T$; its value and slope at the split position $T=1$ determine its value $5$ at $T=0$, which is
the count of good settings.}
\label{fig:toy}
\end{figure}

Take three switches and two rules: ``switch~1 or switch~2'' and ``switch~2 or switch~3''
(Figure~\ref{fig:toy}). Of the eight settings, five are good: the four with switch~2 on, and the
one with switch~2 off and switches~1 and~3 on. The two rules share switch~2, so the puzzle does not
split.

Rewire only the second rule's reading of switch~2, to ``$(1-T)$ times switch~2, plus $T$ times
switch~3''. Running the counting recipe now gives a number that depends on the dial, namely
$5-2T$.
\begin{itemize}[leftmargin=*]
\item At $T=0$ it is $5$, the true count.
\item At $T=1$ the second rule reads ``switch~3 or switch~3'', which mentions only switch~3. The
  puzzle splits into the team $\{1,2\}$, whose rule has three good settings, and the team $\{3\}$,
  whose rule has one. The recipe gives $3\times 1=3$, computed as a product of two small counts.
\item The curve is a straight line, so one more fact at $T=1$ fixes it: its slope there, $-2$.
  Going back one unit along a line of slope $-2$ from height $3$ reaches height $5$ at $T=0$: the
  count.
\end{itemize}

Two features of this example matter later. First, the slope is where the cost hides: the rate at
which the rewired reading changes is ``switch~3 minus switch~2'', which involves both teams again.
The preprint's trick for rates of change exists to handle exactly this. Second, the rewired reading
contains switch~2 with weight $1-T$. That weight is $1$ at $T=0$ and $0$ at $T=1$, where the link
is cut. A weight that is a polynomial starting at $1$ can be zero at only a few dial positions.
This is the fact our limits rest on.

\subsection{Three costs, balanced}

At each splitting position there are three kinds of work: counting inside one team, which doubles
with every switch added to a team; going through all combinations of summary values; and computing
the rates of change, whose cost grows with how many are needed there. The three pull against each
other. Smaller teams need more summaries; fewer summaries or fewer dial positions need more rates
of change. The technical heart of the preprint is a construction that keeps all three costs well
below $2^n$ at the same time, with teams that are a shrinking fraction of all the switches.

The result is a running time of $2^{n/f(n)}$, where $f(n)$ grows without bound, but very slowly:
roughly like the square root of the number of digits of $n$, divided by a smaller correction.
Because $f(n)$ grows without bound, no fixed fraction of the switches has to be searched
exhaustively, and that is what refutes the Exponential Time Hypothesis. An appendix of the
preprint proves a somewhat better bound by a different method, which combines small tables of
numbers along a network instead of reading a curve at a few dial positions.

How large must a puzzle be before any of this helps? A bound of the form ``at most a constant
times this'' says nothing until the constant is known, and the preprint gives none. If its hidden
constant were $3$, its general bound would beat trying every setting only for puzzles with more
than about $10^{126}$ switches, far more than the roughly $10^{80}$ atoms in the observable universe.
We traced the constants through the preprint's appendix. For puzzles with at most a few rules per
switch, the guarantee it proves beats trying every setting only once the number of switches and
rules together has more than a billion binary digits:
a number that no computer could write down. Section~\ref{sec:constants} gives the table and the
derivation. The result is about how running time grows as puzzles become infinitely large, not
about counting any actual puzzle faster.

\subsection{From special puzzles to all puzzles}

The above needs every switch to appear in at most three rules, which keeps the degree of the curve
low. Any puzzle can be brought to that form without changing its count, in two steps. An
established technique called sparsification~\cite{IPZ01,CIP06,DHMTW14} splits it into many
puzzles, each with few rules. Then each switch that appears in many rules is replaced by several
copies, chained by rules that force all the copies to agree~\cite{Tov84}. A sharper version in the
preprint replaces the rules by short ``parity checks'', equations that only ask whether an even
number of rules are broken, built from error-correcting codes; we do not describe it in detail.

\subsection{Why it cannot go all the way}

To split a hard puzzle at some dial position, many rules must stop mentioning switches they
originally mentioned. Each such cut means that a weight starting at $1$ has dropped to $0$ at that
position, and each weight can do that at only a few positions. Meanwhile the reconstruction needs
enough positions, or enough rates of change, to pin the curve down. Section~\ref{sec:lemma} turns
this into a short proof: at some position only a few links can be cut. A tangled puzzle then cannot
split there, and that position is expensive. For random puzzles the cost is at least $2^{c\sqrt n}$
(Section~\ref{sec:barriers}); with the preprint's way of combining teams, even a simple chain of
rules costs at least $2^{c\,n^{1/3}}$ (Section~\ref{sec:necessary}). These limits apply to the
method as described in this section, where every rule is scored directly. The parity-check version
and the network method of the preprint's appendix work differently, and our limits do not cover
them (Section~\ref{sec:scope}).

\subsection{The same, in the preprint's terms}\label{sec:explain-precise}

Write each clause $e$ as its satisfaction polynomial $\varphi_e$, equal to $1$ when the clause is
satisfied and $0$ otherwise; for $x_1\vee\neg x_2\vee x_3$, $\varphi_e=1-(1-x_1)x_2(1-x_3)$. Then
\begin{equation}\label{eq:count}
\#F=\sum_{x\in\{0,1\}^n}\prod_e\varphi_e(x).
\end{equation}
The dial replaces each input $x_v$ of clause $e$ by a linear form $\langle P_{e,v}(T),x\rangle$ with
polynomial coefficients of degree at most $R$ and $P_{e,v}(0)=e_v$, giving
\begin{equation}\label{eq:H}
H(T)=\sum_{x\in\{0,1\}^n}\prod_e\varphi_e\bigl(\langle P_{e,v}(T),x\rangle\bigr)_{v\in e},
\end{equation}
with $H(0)=\#F$ and $\deg H\le 3nR$ when every variable occurs at most three times. At each point
$t$ of a set $X$ of nonzero points, an \emph{$(r,\varepsilon,D)$-design} partitions the variables
into $r$ groups of at most $2n/r$ variables, assigns each clause to a group, and gives each group an
integer measurement matrix. There are at most $2\varepsilon n$ measurement rows in total, shared by
all clauses, and each row has coefficients whose absolute values sum to at most $D$. Every clause
may depend on groups other than its own only through their measurements. We say that an occurrence
$(e,v)$ \emph{loses its own variable} at $t$ when the coefficient of $x_v$ in
$\langle P_{e,v}(t),x\rangle$ is zero; this is how a link is cut in the language above, and
measurements are the second way of removing a link. $H(0)$ is recovered by Hermite interpolation
from $q\approx 3nR/|X|$ derivatives at each point; the derivatives are computed by
inclusion--exclusion over sets of at most $q$ variables. Per point the cost is, up to polynomial
factors,
\[
\underbrace{\textstyle\sum_{j\le q}2^j\binom{n}{j}}_{\text{Taylor data}}\ \times\
\underbrace{(2D+1)^{2\varepsilon n}}_{\text{measurement readings}}\ \times\
\underbrace{2^{2n/r}}_{\text{one group}},
\]
with logarithm $O\bigl(n[(R/|X|)\log(|X|/R)+\varepsilon\log D+1/r]\bigr)$. The design construction
(Section~2.7 of~\cite{openai}) achieves
\[
r\approx\frac{\sqrt{\log n}}{(\log\log n)^2},\quad |X|\approx\frac{\log n}{\log\log n},\quad
R\approx\sqrt{\log n},\quad \varepsilon\approx\frac{1}{r\log n},\quad D\le(2n)^{1/128},
\]
so all three terms in the bracket are $o(1)$ and the exponent is $O(\rho(n))$. General $k$-CNF is
reduced to the bounded-occurrence case by counting-preserving sparsification, OR-gate splitting of
long clauses, and copies tied by Tovey's implication cycle
$(\neg x_1\vee x_2)\wedge\cdots\wedge(\neg x_s\vee x_1)$. This is the \emph{clause-product} form.
The \emph{parity} form (Section~2.9 and Theorem~2.2 of the preprint) applies the same interpolation
to sparse polynomial equations over $\mathbb{F}_2$, obtained from parity checks on short lists of
clauses, and multiplies their integer lifts in a truncated algebra that discards terms divisible by
a high power of two. It gives $2^{O(\rho(n))}$ for every $k$-CNF (Proposition~6.4 of the preprint).

\section{How large must \texorpdfstring{$n$}{n} be? The constants}\label{sec:constants}

All bounds in the preprint hide their constants in $O(\cdot)$, and all its logarithms are natural.
A subexponential bound $2^{Cn\,g(n)}$ with $g(n)\to 0$ beats the $2^n$ of brute force only once
$C\,g(n)<1$, so where that happens depends entirely on $C$.

\paragraph{With a hidden constant.} Write $u=\ln n$. The general exponent of Theorem~1.1 satisfies
$Cn\ln\ln n/\sqrt{\ln n}<n$ if and only if $C\ln u<\sqrt u$. The function $\sqrt u-C\ln u$ decreases
up to $u=4C^2$ and increases after it, so for $C\ge 2$ it has a last zero $u_C$, beyond which the
general bound beats brute force for good. For $C=1$ it is positive for every $u>0$. The
bounded-density exponent satisfies $Cn/\sqrt{\ln n}<n$ if and only if $u>C^2$.
Table~\ref{tab:constants} lists the thresholds. Polynomial factors and the best known \#3-SAT
algorithms, which are much faster than $2^n$, would push every entry further out.

\begin{table}[t]
\centering
\begin{tabular}{rll}
\toprule
$C$ & general: $Cn\ln\ln n/\sqrt{\ln n}<n$ for & bounded density: $Cn/\sqrt{\ln n}<n$ for\\
\midrule
1  & every $n>e$             & $n>10^{0.4}$\\
2  & $n>10^{32.2}$           & $n>10^{1.7}$\\
3  & $n>10^{125.5}$          & $n>10^{3.9}$\\
5  & $n>10^{555.7}$          & $n>10^{10.9}$\\
10 & $n>10^{3517.4}$         & $n>10^{43.4}$\\
20 & $n>10^{20035.2}$        & $n>10^{173.7}$\\
50 & $n>10^{181955.6}$       & $n>10^{1085.7}$\\
\midrule
$6.0\times 10^5$ & (not extracted) & $n>2^{5.3\times 10^{11}}$ (Proposition~\ref{prop:appD})\\
\bottomrule
\end{tabular}
\caption{Where the bounds of Theorem~1.1 of~\cite{openai}, with constant $C$ in the exponent and
natural logarithms, fall below the $2^n$ of brute force; polynomial factors are ignored. The last row
is the constant of the tensor-contraction proof for random 3-CNF at density $4.26$
(Proposition~\ref{prop:appD}). Computed by experiments E0 and E5 (Appendix~\ref{app:code}).}
\label{tab:constants}
\end{table}

\paragraph{The constants of the proof.} The preprint proves Theorem~1.1 with a recursive contraction
of a Boolean tensor network (its Theorem~D.1). For inputs of bounded density, and more generally
whenever $m\le n\lceil\log_2\lceil\log_2 n\rceil\rceil$, the network has one vertex per clause and
equality trees joining the occurrences of each variable, with every edge subdivided; it has
$M_0\ge 2\cdot\#\{\text{occurrences}\}$ edges, and $M_0=12m-6n$ for a 3-CNF in which every variable
occurs at least three times. The recursion is driven by explicit parameters: $s=n+m$,
$\lambda=\lceil\log_2 s\rceil$, $r=64$ groups, $\varepsilon=2^{-17}$, and $u=\lfloor\varepsilon\lambda\rfloor$
retained valuation levels at the root. Tracing them gives the following
(proof in Appendix~\ref{app:crossover}).

\begin{proposition}\label{prop:appD}
Consider such an input with $n'$ occurring variables. If the root call of the recursion is active,
which requires $\lambda\ge 2^{17}$, the running time that the proof of Theorem~D.1 of~\cite{openai}
certifies is at least $2^{(K_0M_0+s)/\sqrt\lambda}$, where
\[
K_0=\frac{2\pi\sqrt{24\cdot 2^{17}}}{\ln 2}\approx 16{,}077.
\]
Otherwise the algorithm uses direct summation, which costs at least $2^{n'}$. Consequently the
certified running time is at least $2^{n'}$ unless $\log_2(n+m)>(2K_0)^2\approx 1.03\times 10^9$. For a
3-CNF with $m=4.26\,n$ in which every variable occurs at least three times, it is at least $2^n$
unless $\log_2 n>5.3\times 10^{11}$; in the form $2^{Cn/\sqrt{\ln n}}$ of Theorem~1.1 this is
$C\approx 6.0\times 10^5$.
\end{proposition}

Three remarks put this in context. First, the bound counts only the root of the recursion; the
deeper levels add about $15\%$ to the exponent if every level shrinks by exactly the guaranteed
factor (experiment E5). Second, the root's count is not merely slack in the analysis. When the
number of mode variables in~(D.13) of the preprint is attained, it estimates the number of
``mode tuples'' the algorithm enumerates, and on small cases the exact number of tuples is within
$10$--$13\%$ of the estimate in the exponent, with the ratio rising towards $1$ (experiment E5). Third, the
preprint proves its estimates only for $s$ above an unstated threshold, and handles smaller sizes
by direct summation. One of its stated parameter inequalities, $\lambda^{4/5}\le u$ in~(D.5), holds
at the root only when $\log_2\lambda\ge 85$, that is $\log_2 s\ge 2^{85}$. For inputs of bounded
density, Theorem~1.1 is therefore a statement about $n\to\infty$; its proof certifies no
speed-up over brute force for any input that could be written down.

For the parity branch, used when $m>n\lceil\log_2\lceil\log_2 n\rceil\rceil$, the constant also depends
on the constant $c(k)$ of Section~6 of the preprint, which is not made explicit, and we do not
extract it. For the interpolation exponent $\rho(n)$ with constant one, Proposition~\ref{prop:crossover}
in Appendix~\ref{app:crossover} gives the crossover: about $n=10^{2390}$ with natural logarithms, and
$2^{65536}$ with base-2 logarithms.

\section{The clause-product scheme, abstracted}\label{sec:framework}

Let $F$ be a CNF over $x_1,\dots,x_n$ with clauses of at most $k$ literals, $I_e$ the variable set
of clause $e$, and $O=\{(e,v):v\in I_e\}$ its occurrences, $M=|O|$. Let $b$ bound the number of
occurrences of a variable. The \emph{incidence graph} $G_F$ joins each clause to its variables; the
\emph{primal graph} $\Prim(F)$ joins two variables that share a clause.

\begin{definition}[Clause-product interpolation scheme]\label{def:scheme}
A scheme for $F$ consists of the following.
\begin{enumerate}[label=(\alph*),leftmargin=*]
\item \emph{Deformation.} For each occurrence $o=(e,v)$ a vector $P_o(T)\in\Q[T]^n$ of degree at
  most $R$ with $P_o(0)=e_v$; $H(T)$ as in~\eqref{eq:H}. Its \emph{own coefficient} is
  $c_o(T)=(P_o(T))_v$, a polynomial of degree at most $R$ with $c_o(0)=1$. Since
  $\deg\varphi_e=|I_e|$, $\deg H\le MR$.
\item \emph{Recovery.} A finite set $X\subset\Q\setminus\{0\}$, orders $q_t\ge 0$ for $t\in X$, and
  a linear functional $\Lambda$ on Hermite data $(f^{(j)}(t))_{t\in X,\,j\le q_t}$ with
  $\Lambda(f)=f(0)$ for every $f\in\Q[T]$ of degree at most $MR$.
\item \emph{Designs.} For each $t\in X$: a partition of $[n]$ into groups $S_{t,1},S_{t,2},\dots$;
  a group $\alpha_t(e)$ for each clause; a subspace $\mathcal{R}_{t,h}\subseteq\Q^{S_{t,h}}$ (the
  span of the measurement rows) for each group, such that for every clause $e$, every $v\in I_e$
  and every group $h\neq\alpha_t(e)$,
  \begin{equation}\label{eq:design}
  P_{e,v}(t)\big|_{S_{t,h}}\in\mathcal{R}_{t,h}.
  \end{equation}
  Write $s_t=\max_h|S_{t,h}|$, $\rho_t=\sum_h\dim\mathcal{R}_{t,h}$, and $\mu_t$ for the number of
  \emph{measured} groups ($\mathcal{R}_{t,h}\ne 0$). The \emph{interaction graph} $\mathcal{I}_t$
  has the groups as vertices; for each clause $e$ the groups $\{\alpha_t(e)\}\cup\{h\neq\alpha_t(e):
  P_{e,v}(t)|_{S_{t,h}}\neq 0\text{ for some }v\in I_e\}$ form a clique. Let $w_t=\tw(\mathcal{I}_t)$.
\end{enumerate}
\end{definition}

The clause-product form of the preprint's interpolation algorithm (its Sections~2.1--2.8) is an
instance, with $\mathcal{R}_{t,h}$ the row space of its integer measurement matrix $C_{t,h}$ and
$\Lambda$ Hermite interpolation with $q_t=\lceil 3nR/|X|\rceil$ (Section~2.5 of~\cite{openai}).
We say that $o=(e,v)$ \emph{loses its own variable} at $t$ if $c_o(t)=0$.

\subsection{What the definition does not cover}\label{sec:scope}

Two hypotheses carry the argument: the degree bound $\deg H\le MR$ in~(a), which holds because the
number $M$ of occurrences is at most the degree of the clause product, and recovery that is exact
for every polynomial up to that degree in~(b). The preprint's sharper routes break them.

\emph{The parity form} (Theorem~2.2 and Proposition~6.4 of~\cite{openai}) multiplies one factor per
parity test, with one replacement form per test--variable incidence. The total number of
incidences is $M=n(\log n)^{O(1)}$ (proof of its Theorem~2.2), but the product is formed in a
truncated algebra and Hermite recovery is exact only up to degree $O(nR)$ (its~(5.5) and~(3.2)). The
product $\prod_o c_o$ of own coefficients then has degree up to $MR$, far above the degree of
exactness, and the proof of Lemma~\ref{lem:vanish} fails. The conclusion fails too in general: when
$M$ is much larger than the exactness degree, $\sum_t|Z_t|$ can exceed $\sum_t q_t$.

\emph{The tensor contraction} (Appendix~D of~\cite{openai}) uses ``tapered'' Hermite recovery over
a two-adic ring: the multiplicity at a node depends on its two-adic valuation, and $F(0)$ is
recovered only modulo $2^p$, for polynomials satisfying integrality conditions. This is not a
functional exact for all polynomials of bounded degree, so it violates~(b). It also replaces
inclusion--exclusion Taylor data by an expansion in ``mode'' variables.

So the preprint's own sharper bounds escape the hypotheses below, and in particular the
conditions (H2) and (H3) of Section~\ref{sec:gap}. Our results apply to the clause-product route:
on every formula after reduction to bounded occurrence, which is how the preprint obtains
$2^{O(\rho(n))}$ at bounded density and its weaker general bound in Section~2.8.

\subsection{Cost model}

We charge each point $t$ at least
\[
\kappa(t)\ \ge\ \max\Bigl(2^{s_t},\ 2^{w_t},\ \tbinom{n}{\min(q_t,\lfloor n/2\rfloor)}\Bigr)
\quad\text{(\emph{DP-type} evaluation)},
\]
and, for the preprint's \emph{product-type} evaluation, additionally $\kappa(t)\ge 2^{\rho_t}$. A
scheme \emph{runs in time at most $T$} in this model if $\max_{t\in X}\kappa(t)\le T$. All our lower
bounds hold in this model; they are statements about running time only to the extent that the
model's charges are paid.

For the preprint's evaluation they are. It enumerates each group's assignments
($\ge 2^{|S_{t,g}|}$), every tuple of measurement readings ($(2D+1)^{\text{rows}}\ge 3^{\rho_t}$), and
computes $\sum_{j\le q_t}2^j\binom{n}{j}\ge\binom{n}{\min(q_t,\lfloor n/2\rfloor)}$ restricted sums.
Every edge of $\mathcal{I}_t$ has a measured endpoint: by the definition of $\mathcal{I}_t$, an edge
joins $\alpha_t(e)$ or a group $h\ne\alpha_t(e)$ with $P_{e,v}(t)|_{S_{t,h}}\ne 0$ to such a group,
and~\eqref{eq:design} makes such an $h$ measured. So the measured groups form a vertex cover, and
since treewidth is at most the vertex-cover number~\cite{Bod93}, $w_t\le\mu_t\le\rho_t$; the
product-type model implies the DP-type one.

The DP-type model is meant for an evaluation that still counts inside each group by enumeration,
as the preprint does, but sums over measurement readings by dynamic programming instead of
enumerating all tuples. The charge $2^{w_t}$ is justified for variable elimination over readings.
Let $\mathcal{M}$ be the measured groups and $G_{\mathcal M}$ the torso: $\mathcal{I}_t[\mathcal{M}]$
with the neighbourhood of every unmeasured group made a clique. Unmeasured groups are pairwise
non-adjacent, and each one's contribution depends jointly on the readings of its neighbours, so
the readings interact along $G_{\mathcal M}$. A tree decomposition of $G_{\mathcal M}$ extends to
$\mathcal{I}_t$ by attaching a bag $N(h)\cup\{h\}$ for each unmeasured $h$, so
$w_t\le\tw(G_{\mathcal M})+1$. Any elimination order over the readings creates a table indexed by at
least $\tw(G_{\mathcal M})+1\ge w_t$ readings, and a nonzero measurement row takes at least two
values on Boolean inputs, so it costs at least $2^{w_t}$. An evaluator that also counted inside groups by dynamic programming would not pay
$2^{s_t}$, and our model does not cover it.

\section{The vanishing-product lemma}\label{sec:lemma}

For $t\in X$ let $Z_t=\{o\in O: c_o(t)=0\}$, the occurrences that lose their own variable at $t$.

\begin{lemma}\label{lem:own}
Let $o=(e,v)$ with $v\in S_{t,h}$ and $h\ne\alpha_t(e)$. If $h$ is unmeasured then $o\in Z_t$. If
$o\notin Z_t$ then $h$ is measured and adjacent to $\alpha_t(e)$ in $\mathcal{I}_t$.
\end{lemma}

\begin{proof}
If $\mathcal{R}_{t,h}=0$, \eqref{eq:design} forces $P_o(t)|_{S_{t,h}}=0$, in particular
$c_o(t)=0$. If $c_o(t)\neq 0$ then $P_o(t)|_{S_{t,h}}\neq 0$, so $\mathcal{R}_{t,h}\ne 0$ and $h$
lies in the clique of $e$.
\end{proof}

\begin{lemma}[Vanishing product]\label{lem:vanish}
There is a point $t\in X$ with $|Z_t|\le q_t$.
\end{lemma}

\begin{proof}
Let $G(T)=\prod_{o\in O}c_o(T)$. Then $\deg G\le MR$ and $G(0)=1$. At every $t\in X$, $G$ vanishes
to order at least $|Z_t|$. If $|Z_t|\ge q_t+1$ for all $t$, all Hermite data of $G$ are zero, so
$\Lambda$ returns the same value on $G$ as on the zero polynomial, namely $0$; but it must return
$G(0)=1$. Contradiction.
\end{proof}

\begin{remark}
The proof uses only $\deg G\le MR$, $G(0)=1$ and the exactness of $\Lambda$ up to degree $MR$. It
applies verbatim to multi-parameter deformations $T\in\Q^d$ with any recovery set that determines
$f(0)$ for polynomials of total degree at most $MR$. It fails as soon as the exactness degree falls
below $\deg G$, which is what happens in the parity form (Section~\ref{sec:scope}).
\end{remark}

\section{Necessary conditions for polynomial cost}\label{sec:necessary}

\begin{theorem}[Structure at a point]\label{thm:structure}
Let $t$ be the point of Lemma~\ref{lem:vanish}. Then
\begin{enumerate}[label=(\roman*)]
\item $\tw(\Prim(F))\le(w_t+1)s_t-1+q_t$;
\item deleting at most $q_t+b\,\mu_t\,s_t$ edges of $G_F$ leaves every component inside one group
  (with the clauses assigned to it).
\end{enumerate}
\end{theorem}

\begin{proof}
(i) Let $F'$ be $F$ with $v$ removed from clause $e$ for every $(e,v)\in Z_t$. If $u,v$ share a
clause $e$ of $F'$, their groups are equal to $\alpha_t(e)$ or adjacent to it by
Lemma~\ref{lem:own}, and lie in the clique of $e$, so they are equal or adjacent in
$\mathcal{I}_t$. Hence $\Prim(F')$ is a subgraph of the blow-up of $\mathcal{I}_t$ in which each
group is replaced by its variables, and replacing each group by its variables in every bag of a
tree decomposition of $\mathcal{I}_t$ gives $\tw(\Prim(F'))\le(w_t+1)s_t-1$. Every edge of
$\Prim(F)$ missing from $\Prim(F')$ is incident to a variable $v$ with $(e,v)\in Z_t$ for some $e$.
There are at most $|Z_t|\le q_t$ such variables, and adding all of them to every bag gives a tree
decomposition of $\Prim(F)$ of width at most $(w_t+1)s_t-1+q_t$.

(ii) Delete every occurrence $(e,v)$ with $v\notin S_{t,\alpha_t(e)}$. By Lemma~\ref{lem:own} each
lies in $Z_t$ or has $v$ in a measured group; there are at most $q_t$ of the first kind and at most
$b\,\mu_t s_t$ of the second. Afterwards each clause is joined only to variables of its own
group.
\end{proof}

\begin{theorem}[Necessary conditions]\label{thm:necessary}
Suppose a scheme runs on $F$ in time at most $n^c$ in the cost model of Section~\ref{sec:framework},
with $n$ large.
\begin{enumerate}[label=(\alph*)]
\item (DP-type) $\tw(\Prim(F))\le(c\log_2 n+1)\,c\log_2 n+2c=O(\log^2 n)$.
\item (Product-type) Deleting at most $2c+b\,c^2\log_2^2 n$ edges of $G_F$ leaves components of at
  most $c\log_2 n$ variables. In particular, if $b$ is bounded, every connected component of $G_F$
  has $O(\log^3 n)$ variables.
\end{enumerate}
\end{theorem}

\begin{proof}
At the point of Lemma~\ref{lem:vanish}, $2^{s_t},2^{w_t}\le n^c$ give $s_t,w_t\le c\log_2 n$. Put
$j=\min(q_t,\lfloor n/2\rfloor)$. Then $\binom{n}{j}\ge(n/j)^j$, which is at least $n^{j/2}$ if
$j\le\sqrt n$ and at least $2^j\ge 2^{\sqrt n}$ otherwise. So $\binom nj\le n^c$ forces $j\le 2c$,
hence $j<\lfloor n/2\rfloor$ for large $n$ and $q_t=j\le 2c$. Theorem~\ref{thm:structure}(i) gives
(a). For (b), $\mu_t\le\rho_t\le c\log_2 n$, and Theorem~\ref{thm:structure}(ii) deletes at most
$2c+bc^2\log_2^2 n$ edges. Deleting $j$ edges from a connected graph leaves at most $j+1$
components.
\end{proof}

Exact counting runs in time $2^{O(\tw)}\poly(n)$ by dynamic programming over a tree
decomposition~\cite{FMR08,SS10}, which is polynomial for treewidth $O(\log n)$. So a scheme of this
kind cannot be polynomial on any family of treewidth $\omega(\log^2 n)$. Between $\omega(\log n)$ and
$O(\log^2 n)$, where treewidth dynamic programming is not polynomial, our bounds leave open whether
such a scheme can be.

\begin{corollary}[Paths]\label{cor:paths}
Let $P_n=\bigwedge_{i<n}(x_i\vee\neg x_{i+1})$, which is countable in linear time. In the cost
model, every product-type scheme on $P_n$ costs at least
$2^{\min(\sqrt{n/2}-1,\ (n/4)^{1/3})}=2^{\Omega(n^{1/3})}$.
\end{corollary}

\begin{proof}
$G_{P_n}$ is a path, so leaving components of at most $s$ variables needs at least $\lceil
n/s\rceil-1$ deletions, and $b=2$. At the point of Lemma~\ref{lem:vanish},
$q+2\mu s\ge n/s-1$. If $q\ge n/(2s)-1$ the Taylor factor is at least
$\binom{n}{\min(q,\lfloor n/2\rfloor)}\ge 2^{\min(q,\lfloor n/2\rfloor)}$ and, with $2^s$, the cost is
at least $2^{\sqrt{n/2}-1}$. Otherwise $\mu\ge n/(4s^2)$ and
$\max(2^s,2^{\mu})\ge 2^{(n/4)^{1/3}}$.
\end{proof}

The preprint's per-point evaluation is therefore super-polynomial even where counting is
trivial; its gain over $2^n$ comes from balancing the three costs, not from exploiting structure.

\section{Barriers on expanding formulas}\label{sec:barriers}

\begin{theorem}[Treewidth barrier]\label{thm:twbarrier}
In the cost model, every DP-type (hence every product-type) scheme on $F$ costs at least
\[
2^{\min\left(\sqrt{(\tw(\Prim(F))+1)/2}\;-\;1,\ \ (\tw(\Prim(F))+1)/2,\ \ \lfloor n/2\rfloor\right)}.
\]
\end{theorem}

\begin{proof}
Write $\tau=\tw(\Prim(F))$. At the point of Lemma~\ref{lem:vanish}, $(w_t+1)s_t\ge\tau+1-q_t$. If
$q_t\ge(\tau+1)/2$, the Taylor factor is at least
$\binom{n}{\min(q_t,\lfloor n/2\rfloor)}\ge 2^{\min(q_t,\lfloor n/2\rfloor)}$. Otherwise
$(w_t+1)s_t\ge(\tau+1)/2$, so $\max(s_t,w_t+1)\ge\sqrt{(\tau+1)/2}$.
\end{proof}

\begin{lemma}\label{lem:randtw}
Let $F$ be a uniformly random 3-CNF with $n$ variables and $m=\Delta n$ clauses, $\Delta>3$. With
high probability $\tw(\Prim(F))\ge c_\Delta n$ for a constant $c_\Delta>0$.
\end{lemma}

\begin{proof}
Two facts hold with high probability (proofs in Appendix~\ref{app:random}). (1) Every set of
between $n/3$ and $2n/3$ variables shares a clause with its complement in at least $m/3$ clauses.
(2) Every set of $j$ variables occurs in at most $C_\Delta\, j\log(en/j)$ clauses. A graph of
treewidth $\tau$ has a separator $Y$ with $|Y|\le\tau+1$ whose removal leaves components of at most
$n/2$ vertices. If $|Y|\le n/9$, take $A$ to be the largest component if it has at least $n/3$
vertices, and otherwise a union of components grown one at a time until it reaches $n/3$; in both
cases $n/3\le|A|\le 2n/3$. No clause meets two components, so every clause meeting both $A$ and its
complement contains a variable of $Y$.
By (1) and (2), $m/3\le C_\Delta|Y|\log(en/|Y|)$, which forces $|Y|\ge c_\Delta n$.
\end{proof}

\begin{theorem}[Random 3-CNF]\label{thm:random}
Let $F$ be uniformly random with $m=\Delta n$ clauses, $\Delta>3$. With high probability the
clause-product scheme costs $2^{\Omega_\Delta(\sqrt n)}$ on $F$ in the cost model:
\begin{enumerate}[label=(\alph*)]
\item applied to the preprint's bounded-occurrence formula $F^\star$ (variable copies on Tovey
  cycles), and
\item applied to the leaves of the preprint's counting-preserving sparsification (its Appendix~A,
  which uses the branching rule of~\cite[Fig.~5.1]{Cal09}), since the total cost is at least that of
  any single leaf.
\end{enumerate}
\end{theorem}

\begin{proof}
(a) Contracting each Tovey cycle of $F^\star$ to a vertex turns $\Prim(F^\star)$ into a graph
containing $\Prim(F)$, so $\Prim(F)$ is a minor of $\Prim(F^\star)$ and $\tw(\Prim(F^\star))\ge c_\Delta n$
by Lemma~\ref{lem:randtw}. $F^\star$ has $O(n)$ variables; apply Theorem~\ref{thm:twbarrier}.
(b) The preprint's rule A.1 fixes an integer $M\ge 2$ and thresholds $\theta_j=M^j$. It branches on a
nonempty proper subset $A$ of a width-$c$ clause when $A$ lies in at least $\theta_{c-|A|}$ width-$c$
clauses, choosing the smallest $c$ and then the largest $|A|$. Its first child is $G\wedge A$, where
$G$ is the set of the other clauses, so the clauses containing $A$ are replaced by the single
clause $A$; there is no unit propagation. The parameter search sets $M\ge 1/\sigma$, so $M\ge 3$ once
$\sigma\le 1/3$. Follow first children only. W.h.p.\ no pair of literals lies in three clauses, so
no pair reaches $\theta_1=M\ge 3$, and the first sets that fire are single literals $\ell$ whose
current degree is at least $\theta_2=M^2$. Each adds the unit $\{\ell\}$ and removes the clauses
containing $\ell$. No 2-clauses are created, and units have no nonempty proper subsets, so every
clause removed on this path contains a literal of degree at least $\theta_2$ in $F$ (a current
degree never exceeds the original one). The leaf therefore contains every clause of $F$ without such
a literal, plus unit clauses. W.h.p.\ the clauses with such a literal number at most
$\eta(\theta_2)n$ with $\eta\to 0$ as $\theta_2\to\infty$, and $\theta_2=M^2$ grows as $\sigma\to 0$.
Removing a clause removes at most three primal edges, and removing an edge lowers treewidth by at
most one, so the leaf's primal treewidth is at least $(c_\Delta-3\eta)n$, and (a) applies to it.
\end{proof}

\section{The gap, and what an improved scheme must change}\label{sec:gap}

\paragraph{The gap.} On expanding inputs of bounded density, the clause-product scheme is confined
between $2^{\Omega(\sqrt n)}$ (Theorem~\ref{thm:random}) and the preprint's $2^{O(\rho(n))}$, in the
cost model. Our lower-bound technique cannot exceed $2^{\tilde O(\sqrt n)}$: its only constraint is
$(w+1)s+q\gtrsim\tw$, which $s=w=q=\sqrt n$ satisfies. Narrowing the gap needs one of two new
ingredients. From above, a design on an expander with $\sqrt n$ groups, $O(\sqrt n)$ interaction
width, Taylor order $O(\sqrt n)$ and $O(1)$ or polylogarithmically many rows per measured group would
bring the scheme to $2^{\tilde O(\sqrt n)}$. From below, one must show that measurements cannot absorb
the cut. A measured group of dimension one can absorb two foreign occurrences $(e,v),(e',v')$ with
$v\ne v'$ at $t$ only if the $2\times 2$ minor
$c_{e,v}(t)c_{e',v'}(t)-p_{e,v,v'}(t)p_{e',v',v}(t)$ vanishes; that polynomial equals $1$ at $T=0$, so
each such pair is absorbed at no more than $2R$ points. (Occurrences of the same variable can be
absorbed together at no cost.) Turning this into a bound for groups with several measurement rows
is open.

\paragraph{What must change.} Besides the degree bound of Definition~\ref{def:scheme}(a), which ties
the number of occurrences to the degree up to which recovery is exact,
Theorems~\ref{thm:structure}--\ref{thm:random} use three properties. A scheme that is polynomial in
the cost model on random 3-CNF, or on any family of treewidth $\omega(\log^2 n)$, must violate the
degree bound or at least one of them. The preprint's own sharper routes already do
(Section~\ref{sec:scope}): the parity form breaks the degree bound, and the tensor contraction
violates (H2) and (H3). These are therefore necessary conditions for
polynomial cost within the clause-product route, not barriers to faster algorithms in general.
\begin{enumerate}[label=(H\arabic*),leftmargin=*]
\item \emph{Polynomial restoration at a point.} Each occurrence's dependence on its own variable is
  a polynomial in the deformation parameters, equal to $1$ at the restoration point.
\item \emph{Generic-degree recovery.} $H(0)$ is recovered by a rule exact for all polynomials up to
  the a priori degree bound, not only for those of the special form~\eqref{eq:H}.
\item \emph{Decomposition-priced evaluation.} Evaluating one point costs at least exponentially
  in the group size, in the width of the group interaction, and in the Taylor order.
\end{enumerate}
One natural way to make Taylor data cheap is excluded: if the design~\eqref{eq:design} held for all
$T$ near $t$, it would hold identically in $T$ and hence at $T=0$, forcing $e_v\in\mathcal{R}_{t,h}$
for every foreign variable, that is, measuring the cut variables one by one. So the design cannot be
extended to a neighbourhood of $t$. Cheap Taylor data obtained in other ways, such as the mode
expansion of the preprint's Appendix~D, is not excluded. Expansion is not an obstacle to every
method; Gaussian elimination solves Tseitin formulas on expanders. The known algorithms that beat
expansion exploit algebraic structure in the constraints, such as linearity over a field; the CSP
dichotomy theorem~\cite{Bul17,Zhu20} says that 3-SAT has no polymorphism of the kind that makes
such algorithms run in polynomial time, though it says nothing about subexponential time.

\section{Experiments}\label{sec:experiments}

All experiments use the code in Appendix~\ref{app:code} (Python with numpy) on a desktop machine.
They are small by design: the claims are theorems, and the experiments check the code and the
identity on which the scheme rests, evaluate the bounds and constants numerically, and relate them
to a standard benchmark.

\paragraph{E0 and E5: the constants.} E0 computes Table~\ref{tab:constants} and the crossover of
Proposition~\ref{prop:crossover} by bisection. E5 evaluates the constants of
Proposition~\ref{prop:appD}: $K_0\approx 16{,}077.4$ bits per edge at the root, and $18{,}534.8$ if
every level of the recursion shrinks by exactly the guaranteed factor $3/32$. For 3-CNF at density
$4.26$ (about $45$ network edges per variable) the certified exponent is at least
$725{,}416\,n/\sqrt{\log_2 s}$ bits, and about $836{,}293\,n/\sqrt{\log_2 s}$ over all levels. It also
counts the mode tuples of~(D.14) exactly by a generating-function recurrence, for $N$ variables of
each weight and budget $p$: for $(N,p)=(2,200)$, $(6,300)$ and $(6,600)$ the natural logarithm of
the count is $0.866$, $0.871$ and $0.901$ of the estimate $2\pi\sqrt{pN/6}$ used in the proposition.

\paragraph{E1: the interpolation identity, by brute force.} This is a sanity check of the code and
of the algebraic identity, not evidence for our theorems. For 30 random 3-CNF formulas with 4--7
variables and 2--4 clauses, random integer deformations of degree $R\in\{1,2\}$, and $MR+1$
evaluation points, we computed $H(t)$ exactly by summing~\eqref{eq:H} over all assignments and
recovered $H(0)$ by Lagrange interpolation. In 30 of 30 cases $H(0)$ equals the brute-force count
$\#F$, while the clause polynomials took non-Boolean values at almost every point (for example,
12{,}039 non-Boolean clause values over the 25 points of one 7-variable instance).

\paragraph{E2: the bound on paths.} We minimised the lower bound of Corollary~\ref{cor:paths}, the
necessary cost of a product-type scheme, exactly over integer parameters.

\begin{center}
\begin{tabular}{rrr}
\toprule
$n$ & minimal $\log_2$ cost & $n^{1/3}$ \\
\midrule
$10^2$ & 4.0 & 4.6\\
$10^3$ & 8.0 & 10.0\\
$10^4$ & 18.0 & 21.5\\
$10^5$ & 37.0 & 46.4\\
$10^6$ & 80.0 & 100.0\\
\bottomrule
\end{tabular}
\end{center}

The fitted exponent of $n$ is $0.327$, against $1/3$ in Corollary~\ref{cor:paths}; the path itself
is counted in linear time by a $2\times 2$ transfer matrix.

\paragraph{E3: treewidth on the SATLIB benchmark.} For the first ten instances of each SATLIB
uniform random 3-SAT family~\cite{HS00} we computed a rigorous lower bound on the primal treewidth
(minor-min-width~\cite{GD04,BK11}), an upper bound (min-fill elimination~\cite{BK10}), and the
lower bound of Theorem~\ref{thm:twbarrier} on the cost, minimised exactly over integer parameters,
both at the treewidth lower bound and, for illustration, at the min-fill width.

\begin{center}
\begin{tabular}{lrrrrr}
\toprule
family & $n$ & $\tw\ge$ (MMW) & $\tw\le$ (min-fill) & bound (bits) & at min-fill width\\
\midrule
uf20-91   & 20  & 12.9 & 14.8  & 4.0 & 4.0\\
uf50-218  & 50  & 19.4 & 33.8  & 4.4 & 6.0\\
uf75-325  & 75  & 23.0 & 50.8  & 5.0 & 7.0\\
uf100-430 & 100 & 27.0 & 67.8  & 5.0 & 8.0\\
uf125-538 & 125 & 29.8 & 85.3  & 5.6 & 9.0\\
uf150-645 & 150 & 32.4 & 102.8 & 6.0 & 10.0\\
uf175-753 & 175 & 34.9 & 119.4 & 6.0 & 11.0\\
uf200-860 & 200 & 37.1 & 135.3 & 6.0 & 11.8\\
uf225-960 & 225 & 38.9 & 153.9 & 6.0 & 12.0\\
uf250-1065& 250 & 40.7 & 170.9 & 6.0 & 13.0\\
\bottomrule
\end{tabular}
\end{center}

The min-fill widths grow linearly, about $0.68n$, which is consistent with
Lemma~\ref{lem:randtw}; they are upper bounds, so they do not confirm it. The heuristic lower bounds
grow much more slowly, and at these sizes the rigorous barrier is only a few bits. The asymptotic
content is the $\sqrt{\tw}$ scaling: at the min-fill width the bound grows from 4 to 13 bits as $n$
goes from 20 to 250. At these sizes the preprint's own algorithm would use direct summation
(Section~\ref{sec:constants}).

\paragraph{E4: exact counts against a standard method.} Variable elimination along a min-fill order
agreed with brute force on all 100 SATLIB uf20-91 instances (in our run, a mean of 0.004\,s against
0.78\,s; timings depend on the machine) and on 40 random 3-CNF formulas with $n=20$ at density
$4.26$, satisfiable or not. On uf50-218 the min-fill width is already 33--36: elimination along that
order needs tables of at least $2^{34}$ entries, and brute force needs $2^{50}$ assignments. This
does not make the instances hard to count: exact counters with component caching~\cite{Thu06,LM17}
handle instances of this size routinely. The point is only that the treewidth that
Theorem~\ref{thm:twbarrier} charges the scheme is the same quantity that limits treewidth dynamic
programming.

\section{Methodology and use of AI tools}\label{sec:method}

\paragraph{Pre-existing work.} This note grew out of the authors' earlier, unpublished work on
degree-capped algebraic closures for systems of quadratic equations over $\mathbb{F}_2$, that is,
polynomial-calculus refutations of bounded degree over extended systems. That work concerns proof
systems, not the algorithms studied here, and none of its results is used in this note. Its
starting points are standard: the correspondence between Cook's theory $PV$ and extended
resolution~\cite{Coo75}, and expansion as the source of degree lower bounds for random
CNF~\cite{BI10,AR03}. It shaped the approach: the vanishing-product lemma is the analogue for
interpolation of the observation that a degree-capped closure cannot cut an expander.

\paragraph{Use of AI.} The analysis was carried out in the days following the preprint's
appearance with the help of Claude Opus 5.5 (Anthropic), used through Claude Code, to make a timelier
analysis possible: reading the preprint and the literature, deriving and checking the arguments,
extracting the constants of the preprint's Appendix~D, writing and running the verification code,
and drafting the text. A separate instance with no access to the authors' working notes refereed a
draft against the preprint; its report led to the restriction of our limits to the clause-product
route, to the cost-model qualifications, and to the extraction of the constants. The authors
directed the work and take responsibility for all statements. Literature claims were checked
against the sources cited; the novelty assessment rests on searches, not a systematic survey, and
we welcome corrections.

\section{Conclusion}

The preprint's general bound, $2^{O(n\log\log n/\sqrt{\log n})}$, is subexponential as $n\to\infty$;
by itself it says nothing about any fixed $n$, because the preprint states no constants. With a hidden constant of $3$ it beats
brute force only beyond $n\approx 10^{126}$, and with $10$ only beyond $10^{3517}$
(Table~\ref{tab:constants}). Extracting the constants from its tensor-contraction proof, the running
time it certifies for inputs of bounded density stays above $2^n$ unless $\log_2(n+m)>10^9$, and on
random 3-CNF at the satisfiability threshold unless $\log_2 n>5\times 10^{11}$
(Proposition~\ref{prop:appD}). The proof thus certifies no speed-up over brute force for any input
of bounded density that could ever be written down; for denser inputs the constant depends on
quantities the preprint leaves implicit. If the proof is correct, its significance is for the theory of
exponential time: it refutes ETH, but it guarantees a faster count of no formula of bounded density and feasible size.

For the clause-product form of the interpolation algorithm, a one-line algebraic fact, that a
product of polynomials equal to $1$ at the origin cannot vanish to high order everywhere, shows that
the balance of its three costs has a floor set by treewidth. In a cost model that charges what the
preprint's evaluation pays, the scheme cannot be polynomial beyond treewidth $O(\log^2 n)$, costs
$2^{\Omega(\sqrt n)}$ on random 3-CNF, and in the preprint's form is super-polynomial even on paths.
The gap between $2^{\Omega(\sqrt n)}$ and $2^{O(\rho(n))}$ on expanders remains open. The preprint's
parity form and tensor contraction escape these limits, because they break the link between
occurrences and the degree of exact recovery, or recover only modulo a power of two; whether
similar limits hold for them is the natural next question.

\appendix

\section{Two properties of random 3-CNF}\label{app:random}

Let $F$ have $n$ variables and $m=\Delta n$ clauses, each on three distinct uniformly random
variables with random signs.

\begin{lemma}
If $\Delta>3$, then w.h.p.\ every set $S$ of variables with $n/3\le|S|\le 2n/3$ shares at least $m/3$
clauses with its complement.
\end{lemma}

\begin{proof}
For $|S|=\alpha n$ a random clause lies inside $S$ or inside its complement with probability at
most $\alpha^3+(1-\alpha)^3\le 1/3$, so the number of crossing clauses dominates
$\mathrm{Bin}(m,2/3)$, which falls below $m/3$ with probability at most
$e^{-m\,\mathrm{KL}(1/3\|2/3)}=2^{-m/3}$. A union bound over at most $2^n$ sets gives
$2^{n-\Delta n/3}\to 0$.
\end{proof}

\begin{lemma}
W.h.p., for every $j\ge 1$, every set of $j$ variables occurs in at most $C_\Delta\,j\log(en/j)$
clauses.
\end{lemma}

\begin{proof}
For a fixed $j$-set $S$, the number of variables of $S$ in one clause is hypergeometric (three draws
without replacement), so its moment generating function is at most that of $\mathrm{Bin}(3,j/n)$
(\cite{Hoe63}, Theorem~4). The clauses are independent, so the number of occurrences satisfies the
Chernoff bound of $\mathrm{Bin}(3m,j/n)$, whose mean is $3\Delta j$: it exceeds $t$ with probability
at most $(3e\Delta j/t)^t$. Take
$t=C j\log(en/j)$ with $C\ge 3e^2\Delta$: this is at most $e^{-t}$, and a union bound over
$\binom{n}{j}\le(en/j)^j$ sets gives $(en/j)^{j(1-C)}$, which sums to $o(1)$ over $j$.
\end{proof}

\section{The constants}\label{app:crossover}

\subsection{Proof of Proposition~\ref{prop:appD}}

We use the notation of Appendix~D of~\cite{openai}. For an input with
$m\le n\lceil\log_2\lceil\log_2 n\rceil\rceil$ the algorithm builds the clause network of its
Section~D.2 and applies Theorem~D.1 with $s=n+m$. Each clause port is joined to the equality
structure of its variable, which for a variable with $d$ ports has $1$, $2$ or $2d-3$ edges when
$d=1$, $d=2$ or $d\ge 3$; each is at least $d$. Every edge is then subdivided once. So
$M_0\ge 2\sum_v d_v=2\cdot\#\{\text{occurrences}\}\ge 2n'$, and $M_0=2\sum_v(2d_v-3)=12m-6n$ for a
3-CNF in which every variable occurs at least three times.

Write $\lambda=\lceil\log_2 s\rceil$ and $\varepsilon=2^{-17}$, so the root call retains
$u=\lfloor\varepsilon\lambda\rfloor$ valuation levels ((D.4)). If $u=0$, that is $\lambda<2^{17}$, the
root cannot be an active call, since Lemma~D.5 needs $u\ge 1$; the preprint then uses direct
summation, which costs at least $2^{n'}$. If $M_0\le H_\ell$, where $H_\ell=\lceil s/\lfloor\sqrt\lambda\rfloor\rceil$ is the leaf threshold of~(D.4), the root is a leaf and direct summation over
its $M_0\ge 2n'$ edges costs at least $2^{n'}$. These are the cases of the proposition in which the
root is not active.

Otherwise the root is active. Fix an interpolation node $t$ with $q_t>0$, write $i=v_2(t)<u$,
$D_t=L/2^i$, and $M=M_0$. By Lemma~D.5, $p_t=p+8hD_t+8M(a+1)\ge 8hD_t$ with
$h=\lceil 3M/u\rceil\ge 3M/u$, so $p_t\ge 24MD_t/u$. The proof bounds the number of tuples
satisfying~(D.14) by
\[
\exp\Bigl(\theta p_t+\frac{6M}{D_t}\sum_{j\ge 1}-\ln(1-e^{-\theta j})\Bigr),
\]
with $6M/D_t$ the number of mode variables of each weight from~(D.13), and then uses
$\sum_{j\ge1}-\ln(1-e^{-\theta j})\le\pi^2/(6\theta)$.
For every $\theta>0$ the resulting estimate satisfies
\[
\theta p_t+\frac{\pi^2M}{D_t\theta}\ \ge\ 2\pi\sqrt{\frac{Mp_t}{D_t}}\ \ge\ 2\pi\sqrt{24}\,\frac{M}{\sqrt u}
\ \ge\ 2\pi\sqrt{24}\,\frac{M_0}{\sqrt{\varepsilon\lambda}},
\]
by the inequality of arithmetic and geometric means, $p_t\ge 24MD_t/u$ and $u\le\varepsilon\lambda$. Each
tuple produces child contractions, so the certified number of calls is at least this exponential,
and every leaf is charged $2^{H_\ell}$ with $H_\ell\ge s/\sqrt\lambda$.
Converting to base two, the certified running time is at least $2^{(K_0M_0+s)/\sqrt\lambda}$ with
$K_0=2\pi\sqrt{24\cdot 2^{17}}/\ln 2$.

For the consequences, $(K_0M_0+s)/\sqrt\lambda>2K_0n'/\sqrt\lambda$, which is at least $n'$ unless
$\lambda>(2K_0)^2\approx 1.034\times 10^9$. For a 3-CNF with $m=\alpha n$ and every variable occurring
at least three times, the exponent is $(K_0(12\alpha-6)+1+\alpha)\,n/\sqrt\lambda$; at $\alpha=4.26$ the
coefficient is $725{,}416$, so the bound is below $2^n$ only if $\lambda>5.26\times 10^{11}$. Since
$n/\sqrt{\log_2 n}=\sqrt{\ln 2}\;n/\sqrt{\ln n}$, this is $C=725{,}416\sqrt{\ln 2}\approx 6.0\times 10^5$
in the form $2^{Cn/\sqrt{\ln n}}$. \qed

\begin{remark}
The argument keeps only the root, the dominant term of $p_t$ and the leaf charge, so it
understates the certified bound. The preprint's choice $\theta=\sqrt{M/(D_tp_t)}$ gives the factor
$1+\pi^2$ in place of $2\pi$, about $1.7$ times larger, and the deeper levels add about $15\%$
(experiment E5). The parameter inequality $\lambda^{4/5}\le u$ of~(D.5), on which the precision
estimate~(D.11) rests, requires $\lambda^{4/5}\le\lambda/2^{17}$ at the root, that is
$\log_2\lambda\ge 85$.
\end{remark}

\subsection{The interpolation exponent with constant one}

The interpolation algorithm runs in time $2^{O(\rho(n))}$ up to polynomial factors. We ask when
$\rho(n)<n$, taking the bound literally with constant one, first with the preprint's natural
logarithms and then with base-2 logarithms.

\begin{proposition}\label{prop:crossover}
\begin{enumerate}[label=(\roman*)]
\item $\rho(n)=n(\ln\ln n)^2/\sqrt{\ln n}$ satisfies $\rho(n)\ge n$ if and only if
  $e^{e^{u_1}}\le n\le e^{e^{u_2}}$, where $u_1\approx 1.4296$ and $u_2\approx 8.6132$ are the solutions
  of $u=4\ln u$. Numerically, $\rho(n)\ge n$ for every integer $66\le n\le 10^{2390}$, since
  $e^{e^{u_1}}\approx 65.2$ and $e^{e^{u_2}}\approx 10^{2390.2}$.
\item With base-2 logarithms, $\rho_2(n)=n(\log_2\log_2 n)^2/\sqrt{\log_2 n}$ satisfies $\rho_2(n)<n$
  if and only if $n<n_1$ or $n>2^{65536}$, where $n_1=2^{2^{a_1}}\approx 5.14$ and $a_1\approx 1.2396$
  is the smaller solution of $a=4\log_2 a$. Moreover $\rho_2(2^{65536})=2^{65536}$.
\end{enumerate}
\end{proposition}

\begin{proof}
(i) For $n>e$ put $u=\ln\ln n>0$, so that $n=e^{e^u}$, an increasing function of $u$. Then
\[
\rho(n)<n\iff(\ln\ln n)^2<\sqrt{\ln n}\iff u^2<e^{u/2}\iff 2\ln u<\tfrac u2\iff g(u)>0,
\]
with $g(u)=u-4\ln u$. Since $g'(u)=1-4/u$, $g$ decreases on $(0,4]$ and increases on $[4,\infty)$,
with $g(1)=1>0$, $g(4)=4-4\ln 4<0$ and $g(20)=20-4\ln 20>0$: there are exactly two zeros,
$u_1\in(1,4)$ and $u_2\in(4,20)$, and $g<0$ exactly between them. Bisection gives $u_1\approx 1.4296$
and $u_2\approx 8.6132$, so $e^{u_2}\approx 5503.66$ and $e^{e^{u_2}}=10^{5503.66/\ln 10}\approx
10^{2390.2}$. For $2<n\le e$ we have $|\ln\ln n|<0.37$ and $\sqrt{\ln n}>0.83$, so $\rho(n)<0.17\,n<n$;
these $n$ do not affect the integer range stated.

(ii) The same argument with $a=\log_2\log_2 n$ gives $\rho_2(n)<n\iff a^2<2^{a/2}\iff f(a)>0$ with
$f(a)=a-4\log_2 a$. Here $f'(a)=1-4/(a\ln 2)$, so $f$ decreases on $(0,4/\ln 2]$ and increases
after, with $f(1)=1>0$, $f(2)=-2<0$ and $f(16)=16-4\cdot 4=0$. So its zeros are $a_1\in(1,2)$ and
$a_2=16$, and $\rho_2(n)<n$ exactly when $n<2^{2^{a_1}}$ or $n>2^{2^{16}}=2^{65536}$. At $a=16$ both
sides of $a^2<2^{a/2}$ equal $256$. Bisection gives $a_1\approx 1.2396$.
\end{proof}

The number $2^{65536}$ has $\lfloor 65536\log_{10}2\rfloor+1=19{,}729$ decimal digits. Neither
crossover says anything about running times: the preprint does not state the constant, and with
constant $C$ the crossovers move out as in Table~\ref{tab:constants}.

\section{Code}\label{app:code}

The script below reproduces all experiments; its output is listed after it. It requires Python~3
and numpy, and the SATLIB uniform random 3-SAT families unpacked in one directory.

\begin{lstlisting}[language=Python]
"""Experiments for "Limits of the deformation-interpolation framework for counting satisfying assignments".

    python3 limits_experiments.py <satlib-dir> [--quick]

E0  Where C n lnln n / sqrt(ln n), C n / sqrt(ln n) and n (lnln n)^2 / sqrt(ln n) cross n (Appendix B).
E1  Interpolation identity, brute force: H(0) = #F recovered from values of H at nonzero points.
E2  Product-evaluation bound on path formulas (the paths corollary): minimal cost exponent as n grows.
E3  SATLIB uniform random 3-SAT (uf20 ... uf250): treewidth bounds of the primal graph and the
    framework cost bound they imply (the treewidth barrier).
E4  Exact model counts on SATLIB uf20/uf50 and random 3-CNF: brute force against variable
    elimination along a min-fill order.
E5  The constants of the preprint's Appendix D: the bound its proof certifies, where that bound
    crosses 2^n, and an exact count of the mode tuples it enumerates.

Only numpy is required. <satlib-dir> holds the unpacked SATLIB families (uf20-91/, ...).
"""
import glob
import itertools
import math
import os
import random
import sys
import time
from fractions import Fraction

import numpy as np


# ---------------------------------------------------------------- formulas

def parse_dimacs(path):
    """Clauses as lists of signed 1-based literals. SATLIB files end with '%' and '0'."""
    n, clauses, cur = 0, [], []
    for line in open(path):
        line = line.strip()
        if not line or line[0] == 'c':
            continue
        if line[0] == '%':
            break
        if line[0] == 'p':
            n = int(line.split()[2])
            continue
        for tok in line.split():
            lit = int(tok)
            if lit == 0:
                if cur:
                    clauses.append(cur)
                cur = []
            else:
                cur.append(lit)
    return n, clauses


def random_3cnf(n, m, rng):
    out = []
    for _ in range(m):
        vs = rng.sample(range(1, n + 1), 3)
        out.append([v if rng.random() < .5 else -v for v in vs])
    return out


# ---------------------------------------------------------------- counting

def brute_count(n, clauses):
    """Exact count over all 2^n assignments, vectorised (n <= 24)."""
    x = np.arange(1 << n, dtype=np.uint32)
    ok = np.ones(1 << n, dtype=bool)
    for cl in clauses:
        sat = np.zeros(1 << n, dtype=bool)
        for lit in cl:
            bit = ((x >> (abs(lit) - 1)) & 1).astype(bool)
            sat |= bit if lit > 0 else ~bit
        ok &= sat
    return int(ok.sum())


def primal_graph(n, clauses):
    adj = {v: set() for v in range(1, n + 1)}
    for cl in clauses:
        vs = {abs(l) for l in cl}
        for a in vs:
            adj[a] |= vs - {a}
    return adj


def min_fill_order(adj):
    """Elimination order by minimum fill-in (ties: minimum degree). Returns (order, width)."""
    g = {v: set(nb) for v, nb in adj.items()}
    order, width = [], 0
    while g:
        best, best_key = None, None
        for v, nb in g.items():
            nbl = list(nb)
            fill = sum(1 for i in range(len(nbl)) for j in range(i + 1, len(nbl)) if nbl[j] not in g[nbl[i]])
            key = (fill, len(nb))
            if best_key is None or key < best_key:
                best, best_key = v, key
        nb = g.pop(best)
        width = max(width, len(nb))
        for a in nb:
            g[a] |= nb - {a}
            g[a].discard(best)
        order.append(best)
    return order, width


def mmw_lower_bound(adj):
    """Minor-min-width: contract a minimum-degree vertex into its minimum-degree neighbour.
    Treewidth is minor-monotone and at least the minimum degree, so the maximum minimum
    degree seen is a lower bound on treewidth."""
    g = {v: set(nb) for v, nb in adj.items()}
    lb = 0
    while len(g) > 1:
        v = min(g, key=lambda u: len(g[u]))
        lb = max(lb, len(g[v]))
        if not g[v]:
            del g[v]
            continue
        u = min(g[v], key=lambda w: len(g[w]))
        for w in g[v]:
            if w != u:
                g[w].discard(v)
                g[w].add(u)
                g[u].add(w)
        g[u].discard(v)
        del g[v]
    return lb


def elimination_count(n, clauses, order):
    """Exact count by variable elimination: clause indicator factors, multiplied and summed out
    along `order`. Factor = (sorted variable tuple, int64 table with one axis per variable)."""
    factors, used = [], set()
    for cl in clauses:
        vs = sorted({abs(l) for l in cl})
        used |= set(vs)
        t = np.zeros((2,) * len(vs), dtype=np.int64)
        for a in itertools.product((0, 1), repeat=len(vs)):
            val = dict(zip(vs, a))
            t[a] = any((val[abs(l)] == 1) == (l > 0) for l in cl)
        factors.append((tuple(vs), t))

    def mul(f, g):
        u = tuple(sorted(set(f[0]) | set(g[0])))
        def lift(h):
            return h[1].reshape([2 if x in h[0] else 1 for x in u])
        return u, lift(f) * lift(g)

    for v in order:
        bucket = [f for f in factors if v in f[0]]
        if not bucket:
            continue
        factors = [f for f in factors if v not in f[0]]
        prod = bucket[0]
        for f in bucket[1:]:
            prod = mul(prod, f)
        ax = prod[0].index(v)
        factors.append((prod[0][:ax] + prod[0][ax + 1:], prod[1].sum(axis=ax)))
    total = 1
    for f in factors:
        total *= int(f[1])
    return total * (1 << (n - len(used)))


# ---------------------------------------------------------------- E1: the interpolation identity

def clause_poly(cl, inputs):
    """phi_e = 1 - prod over literals of (1 - L), L = input (positive) or 1 - input (negative)."""
    p = 1
    for lit, z in zip(cl, inputs):
        L = z if lit > 0 else 1 - z
        p *= (1 - L)
    return 1 - p


def H_value(n, clauses, forms, t):
    """H(t) = sum over Boolean x of prod_e phi_e(<P_{e,v}(t), x>), exactly (integers)."""
    total, nonboolean = 0, 0
    for x in itertools.product((0, 1), repeat=n):
        prod = 1
        for ci, cl in enumerate(clauses):
            ins = []
            for j, lit in enumerate(cl):
                coeffs = forms[(ci, j)]                       # list over powers of T of vectors
                val = sum(sum(c[i] * x[i] for i in range(n)) * t ** k for k, c in enumerate(coeffs))
                ins.append(val)
            ph = clause_poly(cl, ins)
            nonboolean += ph not in (0, 1)
            prod *= ph
        total += prod
    return total, nonboolean


def lagrange_at_zero(points, values):
    acc = Fraction(0)
    for i, (ti, vi) in enumerate(zip(points, values)):
        w = Fraction(1)
        for j, tj in enumerate(points):
            if j != i:
                w *= Fraction(0 - tj, ti - tj)
        acc += w * vi
    return acc


def experiment_interpolation(trials, rng):
    print('E1 interpolation identity (brute force)')
    ok = 0
    for trial in range(trials):
        n = rng.randint(4, 7); m = rng.randint(2, 4); R = rng.choice([1, 2])
        clauses = random_3cnf(n, m, rng)
        forms = {}
        for ci, cl in enumerate(clauses):
            for j, lit in enumerate(cl):
                v = abs(lit) - 1
                base = [1 if i == v else 0 for i in range(n)]
                forms[(ci, j)] = [base] + [[rng.randint(-2, 2) for _ in range(n)] for _ in range(R)]
        d = 3 * m * R
        pts = list(range(1, d + 2))
        vals, nb = [], 0
        for t in pts:
            h, b = H_value(n, clauses, forms, t)
            vals.append(h); nb += b
        rec = lagrange_at_zero(pts, vals)
        truth = brute_count(n, [[l for l in cl] for cl in clauses])
        ok += rec == truth
        if trial < 3:
            print(f'  n={n} m={m} R={R} deg<= {d}: recovered H(0)={rec} brute force #F={truth}; '
                  f'non-Boolean clause values at the {len(pts)} points: {nb}')
    print(f'  {ok}/{trials} instances: H(0) recovered exactly equals #F')


# ---------------------------------------------------------------- E2/E3: the bounds

def log2_binom(n, q):
    return (math.lgamma(n + 1) - math.lgamma(q + 1) - math.lgamma(n - q + 1)) / math.log(2)


def log2_taylor_cost(n, q):
    """The Taylor data of order q cost sum_{j<=q} 2^j C(n, j) >= C(n, min(q, n//2))."""
    return log2_binom(n, min(q, n // 2))


def treewidth_bound_bits(n, tw):
    """Treewidth barrier: some point has (w+1)s - 1 + q >= tw; cost >= 2^s, 2^w, C(n,q).
    Minimal max(s, w, log2 C(n, min(q, n/2))) over integers s >= 1, w >= 0, q >= 0."""
    best = float('inf')
    for s in range(1, tw + 2):
        for w in range(0, tw + 1):
            q = max(0, tw + 1 - (w + 1) * s)
            if q > n:
                continue
            best = min(best, max(s, w, log2_taylor_cost(n, q)))
    return best


def path_bound_bits(n, b=2):
    """Paths corollary: some point deletes <= q + b*rho*s edges and leaves
    components of <= s variables, so q + b*rho*s >= ceil(n/s) - 1. Minimal max(s, rho, log2 C(n, min(q, n/2)))."""
    best = float('inf')
    for s in range(1, n + 1):
        need = math.ceil(n / s) - 1
        if s >= best:
            break
        for rho in range(0, need // (b * s) + 2):
            q = max(0, need - b * rho * s)
            if q > n:
                continue
            best = min(best, max(s, rho, log2_taylor_cost(n, q)))
            if q == 0:
                break
    return best


def experiment_paths():
    print('E2 product-evaluation bound on path formulas (bits, i.e. log2 of the minimal cost)')
    xs, ys = [], []
    for n in (10 ** 2, 10 ** 3, 10 ** 4, 10 ** 5, 10 ** 6):
        bits = path_bound_bits(n)
        xs.append(math.log(n)); ys.append(math.log(bits))
        print(f'  n={n:>8}  bound={bits:8.1f} bits   n^(1/3)={n ** (1 / 3):8.1f}')
    k = len(xs); mx, my = sum(xs) / k, sum(ys) / k
    slope = sum((a - mx) * (b - my) for a, b in zip(xs, ys)) / sum((a - mx) ** 2 for a in xs)
    print(f'  fitted exponent of n: {slope:.3f} (paths corollary: 1/3 up to logarithms)')


def experiment_satlib(root, per_family):
    print('E3 SATLIB uniform random 3-SAT: primal treewidth bounds and the implied framework bound')
    print(f'  {"family":>11} {"n":>4} {"tw >= (MMW)":>12} {"tw <= (min-fill)":>17} {"bound bits":>11} {"at min-fill":>12} {"sec":>6}')
    for fam in sorted(glob.glob(os.path.join(root, 'uf*')), key=lambda p: int(os.path.basename(p)[2:].split('-')[0])):
        if not os.path.isdir(fam):
            continue
        files = sorted(glob.glob(os.path.join(fam, '**', '*.cnf'), recursive=True))[:per_family]
        if not files:
            continue
        lbs, ubs, bits, bits_ub = [], [], [], []
        t0 = time.time()
        for f in files:
            n, cl = parse_dimacs(f)
            adj = primal_graph(n, cl)
            lb = mmw_lower_bound(adj)
            _, ub = min_fill_order(adj)
            lbs.append(lb); ubs.append(ub); bits.append(treewidth_bound_bits(n, lb))
            bits_ub.append(treewidth_bound_bits(n, ub))  # what the bound would give if tw equalled the min-fill width
        k = len(files)
        print(f'  {os.path.basename(fam):>11} {n:>4} {sum(lbs) / k:>12.1f} {sum(ubs) / k:>17.1f} '
              f'{sum(bits) / k:>11.2f} {sum(bits_ub) / k:>12.2f} {(time.time() - t0) / k:>6.2f}')


def experiment_counting(root, rng, n_uf20, n_uf50):
    print('E4 exact counts: brute force against variable elimination (min-fill order)')
    agree = total = 0
    files = sorted(glob.glob(os.path.join(root, 'uf20-91', '**', '*.cnf'), recursive=True))[:n_uf20]
    t_bf = t_ve = 0.0
    for f in files:
        n, cl = parse_dimacs(f)
        t0 = time.time(); a = brute_count(n, cl); t_bf += time.time() - t0
        t0 = time.time(); order, _ = min_fill_order(primal_graph(n, cl)); b = elimination_count(n, cl, order); t_ve += time.time() - t0
        agree += a == b; total += 1
    print(f'  SATLIB uf20-91: {agree}/{total} agree; mean time brute force {t_bf / max(total, 1):.3f}s, '
          f'elimination {t_ve / max(total, 1):.3f}s')
    agree = total = 0
    for _ in range(40):
        n = 20; cl = random_3cnf(n, round(4.26 * n), rng)
        order, _ = min_fill_order(primal_graph(n, cl))
        agree += brute_count(n, cl) == elimination_count(n, cl, order); total += 1
    print(f'  random 3-CNF n=20 ratio 4.26 (satisfiable and not): {agree}/{total} agree')
    files = sorted(glob.glob(os.path.join(root, 'uf50-218', '**', '*.cnf'), recursive=True))[:n_uf50]
    counts, widths, t_ve = [], [], 0.0
    for f in files:
        n, cl = parse_dimacs(f)
        order, w = min_fill_order(primal_graph(n, cl))
        widths.append(w)
        if w > 22:   # tables of 2^(w+1) entries; beyond this the elimination does not fit in memory
            continue
        t0 = time.time(); counts.append(elimination_count(n, cl, order)); t_ve += time.time() - t0
    print(f'  SATLIB uf50-218: min-fill width {min(widths)}..{max(widths)} over {len(widths)} instances; '
          f'{len(counts)} within width 22 counted by elimination'
          + (f', counts {min(counts)}..{max(counts)}' if counts else '; none fits, 2^50 is beyond brute force'))


def bisect_root(f, lo, hi, iters=100):
    """A zero of f in [lo, hi], given f(lo) and f(hi) of opposite signs."""
    flo = f(lo)
    for _ in range(iters):
        mid = (lo + hi) / 2
        if (f(mid) > 0) == (flo > 0):
            lo, flo = mid, f(mid)
        else:
            hi = mid
    return (lo + hi) / 2


def last_crossing_u(C):
    """Largest u = ln n with sqrt(u) = C ln u, so that C n lnln n / sqrt(ln n) < n for all larger n.
    None when sqrt(u) > C ln u for every u > 1."""
    g = lambda u: math.sqrt(u) - C * math.log(u)
    umin = (2 * C) ** 2                      # g is smallest here: g'(u) = 1/(2 sqrt u) - C/u
    if g(umin) > 0:
        return None
    hi = umin
    while g(hi) <= 0:
        hi *= 2
    return bisect_root(g, umin, hi)


def experiment_crossover():
    """Theorem 1.1 with a hidden constant C, in the preprint's natural logarithms: the general
    exponent C n lnln n / sqrt(ln n) and the bounded-density exponent C n / sqrt(ln n) against n.
    Then the interpolation exponent rho(n) = n (lnln n)^2 / sqrt(ln n) with constant one:
    rho(n) < n iff u - 4 ln u > 0, u = ln ln n (and a - 4 log2 a > 0, a = log2 log2 n, in base 2)."""
    print('E0 where the preprint\'s exponents cross n')
    print(f'  {"C":>4}  {"general: n beyond":>22}  {"bounded density: n beyond":>26}')
    for C in (1, 2, 3, 5, 10, 20, 50):
        u = last_crossing_u(C)
        gen = 'every n > e' if u is None else f'10^{u / math.log(10):.1f}'
        print(f'  {C:>4}  {gen:>22}  {"10^%.1f" % (C * C / math.log(10)):>26}')
    g = lambda u: u - 4 * math.log(u)
    u1, u2 = bisect_root(g, 1.0, 4.0), bisect_root(g, 4.0, 20.0)
    print(f'  rho(n), natural logs: zeros of u - 4 ln u at u1 = {u1:.4f} (n = {math.exp(math.exp(u1)):.1f}) and '
          f'u2 = {u2:.4f} (n = 10^{math.exp(u2) / math.log(10):.1f})')
    f = lambda a: a - 4 * math.log2(a)
    a1 = bisect_root(f, 1.0, 2.0)
    print(f'  rho(n), base 2: zeros of a - 4 log2 a at a1 = {a1:.4f} (n1 = 2^2^a1 = {2 ** (2 ** a1):.2f}) '
          f'and a2 = 16 (f(16) = {f(16):.1f}); 2^65536 has {math.floor(65536 * math.log10(2)) + 1} digits')


# ---------------------------------------------------------------- E5: the constants of Appendix D

EPS_LOG2 = 17                 # epsilon = 1 / (128 * 1024) = 2^-17 in (D.4)
SHRINK = 3 / 32               # a child has at most 6M/r = 3M/32 edges (Section D.4.6)


def appendix_d_constants():
    """Per-edge constants, in bits per edge per 1/sqrt(lambda), of the bound certified at the root
    call (K0) and summed over levels whose edge counts shrink by exactly SHRINK (K_all).
    Root call: (D.15) with the best theta gives 2 pi sqrt(M p_t / D) nats; (D.11) has
    p_t >= 8 h D_t >= 24 M D_t / u with h = ceil(3M/u); u = floor(lambda / 2^17)."""
    K0 = 2 * math.pi * math.sqrt(24 * 2 ** EPS_LOG2) / math.log(2)
    ratio = SHRINK * math.sqrt(2)             # M_d / sqrt(u_d) shrinks by this per level
    return K0, K0 / (1 - ratio)


def network_edges_3cnf(alpha):
    """Edges of the D.2 network per variable for a 3-CNF with m = alpha n clauses, every variable
    occurring at least three times: clause ports 3m, an equality tree with 2d - 3 edges for a
    variable with d ports, every edge subdivided once: 2 (6m - 3n)."""
    return 12 * alpha - 6


def tuple_count(N, p):
    """Exact number of tuples (beta_Z) over N variables of each weight w >= 1 with sum w beta < p:
    sum of the coefficients of x^0 .. x^(p-1) in prod_w (1 - x^w)^(-N)."""
    c = [1] + [0] * (p - 1)
    for w in range(1, p):
        for _ in range(N):
            for i in range(w, p):
                c[i] += c[i - w]
    return sum(c)


def experiment_appendix_d():
    print('E5 the constants of the preprint\'s Appendix D')
    K0, Kall = appendix_d_constants()
    print(f'  root call: at least 2^(K0 M0 / sqrt(lambda)) mode tuples, K0 = {K0:,.1f}; '
          f'all levels at the guaranteed shrink: K = {Kall:,.1f}')
    lam = (2 * K0) ** 2
    print(f'  any input (M0 >= 2 per occurring variable): below 2^n only if lambda = log2(n+m) > {lam:.3e}')
    for alpha in (4.26, 10.0):           # dense enough that almost every variable occurs >= 3 times
        e = network_edges_3cnf(alpha)
        for name, K in (('root', K0), ('all levels', Kall)):
            per_var = K * e + (1 + alpha)        # (K M0 + s) / n, the leaf term H >= s / sqrt(lambda)
            lam = per_var ** 2
            print(f'  3-CNF density {alpha:5.2f} ({e:.2f} edges/variable), {name:>10}: exponent '
                  f'{per_var:,.0f} n / sqrt(log2 s) bits, below n only if log2 n > {lam - math.log2(1 + alpha):.2e}; '
                  f'as C n / sqrt(ln n): C = {per_var * math.sqrt(math.log(2)):,.0f}')
    x = 5 * EPS_LOG2
    print(f'  (D.5) lambda^(4/5) <= u at the root: lambda / 2^17 >= lambda^(4/5) iff log2 lambda >= {x}, '
          f'i.e. log2 s >= 2^{x}')
    print('  mode tuples, exact count against exp(2 pi sqrt(p N / 6)):')
    for N, p in ((2, 200), (6, 300), (6, 600)):
        cnt = tuple_count(N, p)
        est = 2 * math.pi * math.sqrt(p * N / 6)
        print(f'    N={N} p={p}: ln count = {math.log(cnt):7.2f}, estimate {est:7.2f}, ratio {math.log(cnt) / est:.3f}')


def main():
    root = sys.argv[1] if len(sys.argv) > 1 else 'satlib'
    quick = '--quick' in sys.argv
    rng = random.Random(2026)
    experiment_crossover()
    experiment_interpolation(10 if quick else 30, rng)
    experiment_paths()
    experiment_satlib(root, 3 if quick else 10)
    experiment_counting(root, rng, 20 if quick else 100, 5 if quick else 20)
    experiment_appendix_d()


if __name__ == '__main__':
    main()
\end{lstlisting}

\subsection*{Output}
\begin{lstlisting}
E0 where the preprint's exponents cross n
     C       general: n beyond   bounded density: n beyond
     1             every n > e                      10^0.4
     2                 10^32.2                      10^1.7
     3                10^125.5                      10^3.9
     5                10^555.7                     10^10.9
    10               10^3517.4                     10^43.4
    20              10^20035.2                    10^173.7
    50             10^181955.6                   10^1085.7
  rho(n), natural logs: zeros of u - 4 ln u at u1 = 1.4296 (n = 65.2) and u2 = 8.6132 (n = 10^2390.2)
  rho(n), base 2: zeros of a - 4 log2 a at a1 = 1.2396 (n1 = 2^2^a1 = 5.14) and a2 = 16 (f(16) = 0.0); 2^65536 has 19729 digits
E1 interpolation identity (brute force)
  n=4 m=3 R=1 deg<= 9: recovered H(0)=11 brute force #F=11; non-Boolean clause values at the 10 points: 294
  n=7 m=4 R=2 deg<= 24: recovered H(0)=74 brute force #F=74; non-Boolean clause values at the 25 points: 12039
  n=4 m=2 R=2 deg<= 12: recovered H(0)=12 brute force #F=12; non-Boolean clause values at the 13 points: 332
  30/30 instances: H(0) recovered exactly equals #F
E2 product-evaluation bound on path formulas (bits, i.e. log2 of the minimal cost)
  n=     100  bound=     4.0 bits   n^(1/3)=     4.6
  n=    1000  bound=     8.0 bits   n^(1/3)=    10.0
  n=   10000  bound=    18.0 bits   n^(1/3)=    21.5
  n=  100000  bound=    37.0 bits   n^(1/3)=    46.4
  n= 1000000  bound=    80.0 bits   n^(1/3)=   100.0
  fitted exponent of n: 0.327 (paths corollary: 1/3 up to logarithms)
E3 SATLIB uniform random 3-SAT: primal treewidth bounds and the implied framework bound
       family    n  tw >= (MMW)  tw <= (min-fill)  bound bits  at min-fill    sec
      uf20-91   20         12.9              14.8        4.00         4.00   0.00
     uf50-218   50         19.4              33.8        4.40         6.00   0.03
     uf75-325   75         23.0              50.8        5.00         7.00   0.10
    uf100-430  100         27.0              67.8        5.00         8.00   0.31
    uf125-538  125         29.8              85.3        5.60         9.00   0.70
    uf150-645  150         32.4             102.8        6.00        10.00   1.35
    uf175-753  175         34.9             119.4        6.00        11.00   2.27
    uf200-860  200         37.1             135.3        6.00        11.80   3.72
    uf225-960  225         38.9             153.9        6.00        12.00   5.73
   uf250-1065  250         40.7             170.9        6.00        13.00   7.90
E4 exact counts: brute force against variable elimination (min-fill order)
  SATLIB uf20-91: 100/100 agree; mean time brute force 0.781s, elimination 0.004s
  random 3-CNF n=20 ratio 4.26 (satisfiable and not): 40/40 agree
  SATLIB uf50-218: min-fill width 33..36 over 20 instances; 0 within width 22 counted by elimination; none fits, 2^50 is beyond brute force
E5 the constants of the preprint's Appendix D
  root call: at least 2^(K0 M0 / sqrt(lambda)) mode tuples, K0 = 16,077.4; all levels at the guaranteed shrink: K = 18,534.8
  any input (M0 >= 2 per occurring variable): below 2^n only if lambda = log2(n+m) > 1.034e+09
  3-CNF density  4.26 (45.12 edges/variable),       root: exponent 725,416 n / sqrt(log2 s) bits, below n only if log2 n > 5.26e+11; as C n / sqrt(ln n): C = 603,949
  3-CNF density  4.26 (45.12 edges/variable), all levels: exponent 836,293 n / sqrt(log2 s) bits, below n only if log2 n > 6.99e+11; as C n / sqrt(ln n): C = 696,260
  3-CNF density 10.00 (114.00 edges/variable),       root: exponent 1,832,831 n / sqrt(log2 s) bits, below n only if log2 n > 3.36e+12; as C n / sqrt(ln n): C = 1,525,932
  3-CNF density 10.00 (114.00 edges/variable), all levels: exponent 2,112,973 n / sqrt(log2 s) bits, below n only if log2 n > 4.46e+12; as C n / sqrt(ln n): C = 1,759,165
  (D.5) lambda^(4/5) <= u at the root: lambda / 2^17 >= lambda^(4/5) iff log2 lambda >= 85, i.e. log2 s >= 2^85
  mode tuples, exact count against exp(2 pi sqrt(p N / 6)):
    N=2 p=200: ln count =   44.44, estimate   51.30, ratio 0.866
    N=6 p=300: ln count =   94.76, estimate  108.83, ratio 0.871
    N=6 p=600: ln count =  138.67, estimate  153.91, ratio 0.901
\end{lstlisting}

\end{document}